\documentclass{emulateapj}

\newcommand{\cha}{\textit{Chandra }}
\def\xmm{{XMM-{\it Newton\/ }}}

\def\leg{{\em COSMOS-Legacy }\/}

\shorttitle{Optical counterparts of the Chandra COSMOS Legacy Survey}
\shortauthors{Marchesi et al.}

\usepackage{float}
\usepackage{color}
\usepackage{graphicx}
\usepackage{gensymb}
\usepackage{enumitem}

\begin{document}

\slugcomment{Accepted to the Astrophysical Journal on November 30, 2015}
\title{The \textit{Chandra} COSMOS Legacy survey: optical/IR identifications}



\author{S. Marchesi\altaffilmark{1,2,3}, F. Civano\altaffilmark{1,2},  M. Elvis\altaffilmark{2}, M. Salvato\altaffilmark{4}, M. Brusa\altaffilmark{3,5}, A. Comastri\altaffilmark{5}, R. Gilli\altaffilmark{5}, G. Hasinger\altaffilmark{6},  G. Lanzuisi\altaffilmark{3,5}, T. Miyaji\altaffilmark{7,8}, E. Treister\altaffilmark{9}, C.M. Urry\altaffilmark{1}, C. Vignali\altaffilmark{3,5}, G. Zamorani\altaffilmark{5}, V. Allevato\altaffilmark{10}, N. Cappelluti\altaffilmark{5}, C. Cardamone\altaffilmark{11}, A. Finoguenov\altaffilmark{10}, R. E. Griffiths\altaffilmark{12}, A. Karim\altaffilmark{13}, C. Laigle\altaffilmark{14},  S. M. LaMassa\altaffilmark{1}, K. Jahnke\altaffilmark{15},  P. Ranalli\altaffilmark{16}, K. Schawinski\altaffilmark{17}, E. Schinnerer\altaffilmark{18,19}, J. D. Silverman\altaffilmark{20}, V. Smolcic\altaffilmark{21}, H. Suh\altaffilmark{6,2}, B. Trakhtenbrot\altaffilmark{17}}

\altaffiltext{1}{Yale Center for Astronomy and Astrophysics, 260 Whitney Avenue, New Haven, CT 06520, USA}
\altaffiltext{2}{Harvard-Smithsonian Center for Astrophysics, 60 Garden Street, Cambridge, MA 02138, USA}
\altaffiltext{3}{Dipartimento di Fisica e Astronomia, Universit\`a di Bologna, viale Berti Pichat 6/2, 40127 Bologna, Italy}
\altaffiltext{4}{Max-Planck-Institut f{\"u}r extraterrestrische Physik, Giessenbachstrasse 1, D-85748 Garching bei M{\"u}nchen, Germany}
\altaffiltext{5}{INAF--Osservatorio Astronomico di Bologna, via Ranzani 1, 40127 Bologna, Italy}
\altaffiltext{6}{Institute for Astronomy, 2680 Woodlawn Drive, University of Hawaii, Honolulu, HI 96822, USA}
\altaffiltext{7}{Instituto de Astronom\'ia sede Ensenada, Universidad Nacional Aut\'onoma de M\'exico, Km. 103, Carret. Tijunana-Ensenada, Ensenada, BC, Mexico}
\altaffiltext{8}{University of California San Diego, Center for Astrophysics and Space Sciences, 9500 Gilman Drive, La Jolla, CA 92093-0424, USA}
\altaffiltext{9}{Universidad de Concepci\'{o}n, Departamento de Astronom\'{\i}a, Casilla 160-C, Concepci\'{o}n, Chile}
\altaffiltext{10}{Department of Physics, University of Helsinki, Gustaf H\"allstr\"omin katu 2a, FI-00014 Helsinki, Finland}
\altaffiltext{11}{Department of Science, Wheelock College, Boston, MA 02215, USA}
\altaffiltext{12}{Physics \& Astronomy Dept., Natural Sciences Division, University of Hawaii at Hilo, 200 W. Kawili St., Hilo, HI 96720, USA}
\altaffiltext{13}{Argelander-Institut f\"ur Astronomie, Universit\"at Bonn, Auf dem H\"ugel 71, D-53121 Bonn, Germany}
\altaffiltext{14}{Sorbonne Universit\'{e}, UPMC Univ Paris 06, et CNRS, UMR 7095, IAP, 98b bd Arago, F-75014, Paris, France}
\altaffiltext{15}{Max Planck Institute for Astronomy, Koenigstuhl 17, D-69117 Heidelberg, Germany}
\altaffiltext{16}{Lund Observatory, P.O. Box 43, 22100 Lund, Sweden}
\altaffiltext{17}{Institute for Astronomy, Department of Physics, ETH Zurich, Wolfgang-Pauli-Strasse 27, CH-8093 Zurich, Switzerland}
\altaffiltext{18}{Max Planck Institute for Astronomy, Königstuhl 17, 69117 Heidelberg, Germany}
\altaffiltext{19}{NRAO, 1003 Lopezville Road, Socorro, NM 87801, USA}
\altaffiltext{20}{Kavli Institute for the Physics and Mathematics of the Universe (WPI), The University of Tokyo Institutes for Advanced Study, The University of Tokyo, Kashiwa, Chiba 277-8583, Japan}
\altaffiltext{21}{Department of Physics, University of Zagreb, Bijeni\v{c}ka cesta 32, HR-10000 Zagreb, Croatia}


\begin{abstract}
We present the catalog of optical and infrared counterparts of the \cha \leg Survey, a 4.6 Ms \cha program on the 2.2 deg$^2$ of the COSMOS field, combination of 56 new overlapping observations obtained in Cycle 14 with the previous C-COSMOS survey. In this Paper we report the $i$, $K$, and 3.6 $\mu$m identifications of the 2273 X-ray point sources detected in the new Cycle 14 observations. We use the likelihood ratio technique to derive the association of optical/infrared (IR) counterparts for 97\% of the X-ray sources. We also update the information for the 1743 sources detected in C-COSMOS, using new $K$ and 3.6 $\mu$m information not available when the C-COSMOS analysis was performed. The final catalog contains 4016 X-ray sources, 97\% of which have an optical/IR counterpart and a photometric redshift, while $\simeq$54\% of the sources have a spectroscopic redshift. The full catalog, including spectroscopic and photometric redshifts and optical and X-ray properties described here in detail, is available online. We study several X-ray to optical (X/O) properties: with our large statistics we put better constraints on the X/O flux ratio locus, finding a shift towards faint optical magnitudes in both soft and hard X-ray band. We confirm the existence of a correlation between X/O and the the 2-10 keV luminosity for Type 2 sources. We extend to low luminosities the analysis of the correlation between the fraction of obscured AGN and the hard band luminosity, finding a different behavior between the optically and X-ray classified obscured fraction.
\end{abstract}



\section{Introduction}
It is widely believed that galaxies and their central supermassive black holes (SMBH) undergo closely coupled evolution. SMBH masses in the nuclei of nearby galaxies correlate with bulge luminosity and stellar velocity dispersion, with a very small scatter (Magorrian et al. 1998; Gebhardt et al. 2000; Ferrarese \& Merrit 2000; Merloni et al. 2010; McConnell \& Ma 2013). Most SMBH - and definitely the most massive ones - had to grow during an active accretion phase, when they would be visible as an active galactic nucleus (AGN), which implies that most bulges had an active phase in their past. Galaxies and AGN show also coeval cosmic ``downsizing'': more luminous AGN and more massive galaxies formed earlier (and therefore their number density peaks at higher redshift) than less luminous AGN and less massive galaxies (Cowie et al. 1996). Massive galaxies exhibit a peak in star formation at $z$$\simeq$2 (Cimatti et al. 2006; Madau and Dickinson 2014), and SMBH growth peaks in the same redshift range ($z$=2-3), as the quasar luminosity function (Hasinger et al. 2005; Silverman et al. 2008; Hasinger 2008; Ueda et al. 2014; Aird et al. 2015; Miyaji et al. 2015). However, even if this common growth seems securely established, the causes of this trend remain largely not understood (e.g. Merloni \& Heinz 2008; Alexander \& Hickox 2012).

The co-evolution of SMBH and galaxies can be studied with sizable samples of AGN, both obscured and unobscured, with sufficient multiwavelength data to disentangle selection effects. To access the moderate luminosity AGN that dominate the X-ray background requires a deep moderate-area survey ($\geq$ 1 deg$^2$ in area, at sufficient depth to detect AGN up to $z\sim$6), on areas wide enough to measure large-scale structures and find rare objects. Moreover, spectroscopic information deep enough to detect faint sources (with L$^*$ luminosities) even at $z\simeq$3 is also required.

X-ray data play an important role in the selection of AGN, because at these energies the contamination from non-nuclear emission, mainly due to star-formation processes, is far less significant than in optical and infrared surveys (Donley et al. 2008, 2012; Stern et al. 2012; Lehmer et al. 2012). X-ray surveys with \cha and \xmm are also very effective at selecting both unobscured and obscured AGN, including also a fraction of AGN in the Compton thick regime, i.e., with hydrogen column densities, $N_H$, up to 10$^{24}$ cm$^{-2}$ and up to $z\simeq$1-2 (Comastri et al. 2011; Georgantopoulos et al. 2013; Lanzuisi et al. 2015; Buchner et al. 2015). Recent works with the hard X-ray telescope \textit{NuSTAR} detected candidate sources even above the 10$^{24}$ cm$^{-2}$ threshold, albeit typically at lower redshift (Lansbury et al. 2014;  Civano et al. 2015a). For these reasons, the combination of X-ray-selected samples of AGN and multiwavelength data is essential to study the evolving properties of accreting SMBHs and their host galaxies. 

In the last 15 years, both \cha and \xmm satellites have been used to survey both deep and wide-area fields (see Brandt \& Alexander 2015 for a review). These surveys produced catalogs of X-ray emitting AGN, which have then been combined with extended multiwavelength spectroscopic and photometric information.

These contiguous surveys follow a ``wedding cake'' strategy, being layered in decreasing area and increasing depth (see Figure 16 in Civano et al. 2015b, hereafter Paper I), to obtain roughly similar numbers of detected sources spanning a broad range in redshift-luminosity space.  At one extreme of this layer are wide/shallow surveys like XBootes (9 deg$^2$; Murray et al. 2005), Stripe 82X (31.2 deg$^{2}$, LaMassa et al. 2013 and submitted), XXL (50 deg$^{2}$, Pierre et al. submitted)  and 3XMM ($\simeq$880 deg$^{2}$, Rosen et al. submitted), which are designed to cover a large volume of the Universe and thus find rare sources, i.e., high-luminosity and/or high-redshift AGN. At the opposite extreme are narrow/ultra-deep surveys like the 4 Ms Chandra Deep Field South (CDF-S, 0.1 deg$^2$; Xue et al. 2011; other 3 Ms of \cha time have already been granted in \cha Cycle 15), which can detect non-active galaxies, even at medium to high redshifts (Luo et al. 2011; Lehmer et al. 2012), and AGN at z$>$5 down to very faint limits, but have statistically small samples of sources at any redshift (e.g., Weigel et al. 2015 showed that the CDF-S does not appear to contain any AGN at z$>$5). 

The \cha \leg project, i.e., the combination of the 1.8 Ms C-COSMOS survey (Elvis et al 2009) with 2.8 Ms of new \cha ACIS-I observations  (Paper I) is exploring a new region of the area versus flux space, by using an unusually large total exposure time (4.6 Ms total) with respect to the observed area (2.15 deg$^2$). 
\cha \leg is deep enough (flux limit $f$$\simeq$2 $\times$ 10$^{-16}$ erg s$^{-1}$cm$^{-2}$ in the 0.5-2 keV band) to find obscured sources with no clear AGN signatures in the optical spectra or spectral energy distributions (SEDs) up to redshift $z\simeq$6 and L$_X\simeq$10$^{45}$ erg s$^{-1}$; at the same time it is wide enough to have one of the largest samples of X-ray point-like sources (4016). Moreover, the \cha \leg sources are also bright enough to obtain almost complete optical and near-infrared (near-IR) identifications of the X-ray sources (97$\%$ in C-COSMOS, Civano et al. 2012b, C12 hereafter): this extended follow-up is also due to the comprehensive nature of the Cosmic Evolutionary Survey (COSMOS; Scoville et al. 2007) and to its multiwavelength photometric and spectroscopic database (Schinnerer et al. 2007; Sanders et al. 2007; Taniguchi et al. 2007; Capak et al. 2007; Koekemoer et al. 2007; Zamojski et al. 2007; Lilly et al. 2007; Trump et al. 2007; Ilbert et al. 2009; McCracken et al. 2010; Laigle et al. submitted). 

The whole COSMOS field was covered previously in the X-rays with XMM-COSMOS (Hasinger et al. 2007; Cappelluti et al. 2009). Therefore, the high luminosity regime of the \cha \leg survey  (L$_X>$10$^{44}$ erg s$^{-1}$) has been already explored in a series of publication from XMM-COSMOS (e.g.  Brusa et al. 2009, 2010, hereafter B10; Allevato et al. 2011; Mainieri et al. 2011; Bongiorno et al. 2012; Lusso et al. 2012, 2013; Merloni et al. 2014; Miyaji et al. 2015 among others). The \cha  low background allows one to reach
fluxes three times fainter in both the 0.5-2 keV and 2-10 keV bands. The combination of area and sensitivities permits to study faint and/or rare systems (e.g. Fiore et al. 2009; Civano et al. 2010, 2012a; Capak et al. 2011) and to measure large-scale clustering in the universe (Allevato et al. 2014). Moreover, \cha can resolve sources with subarcsecond accuracy (Civano et al. 2012b; Lackner et al. 2014).

In this Paper, we present the catalog of optical and infrared counterparts of new \cha \leg sources, presented in Paper I, and we describe and analyze several X-ray and optical/IR photometric and spectroscopic properties of the sources in the whole survey (i.e., combining new and C-COSMOS sources). The Paper is organized as follows: in Section \ref{sec:datasets} we describe the X-ray, optical and infrared catalogs used in this work, in Section \ref{sec:method} we describe the cross-catalog identification technique, while in Section \ref{sec:results} we show the results obtained in the identification process, and in Section \ref{sec:id_prop} we show some basic properties of the different types of optical counterparts. In Section \ref{sec:z_prop} the spectroscopic and photometric redshifts of the survey are described, together with the spectral and SED-based classification, in Section \ref{sec:catalog} we describe the identification catalog, in Section \ref{sec:opt_ir} we analyze the relations between X-ray and optical/IR properties and in Section \ref{sec:conclusions} we summarize the main results of this Paper.

We assume a cosmology with $H_0$= 70 km s$^{-1}$ Mpc$^{-1}$, $\Omega_M$=0.29 and $\Omega_\Lambda$=0.71. The AB magnitude system is used in this Paper if not otherwise stated.

\section{Identification datasets}\label{sec:datasets} 
The X-ray catalog used in this work is obtained from the \textit{Chandra} COSMOS Legacy survey, which properties are extensively described in Paper I. In this Section, we refer to the subsample of  the catalog which contains 2273 new point-like X-ray sources, not previously detected in C-COSMOS, detected down to a maximum likelihood threshold DET\_ML=10.8 in at least one band (0.5-2, 2-7 or 0.5-7 keV), corresponding to a Poisson probability of $P$$\simeq$5$\times$10$^{-5}$ that a detected source is actually a background fluctuation. 
The flux limits of the survey at 20\% of the area of the whole survey are 1.3$\times$10$^{-15}$ erg s$^{-1}$cm$^{-2}$ in the full band (0.5-10 keV), 3.2$\times$10$^{-16}$ erg s$^{-1}$cm$^{-2}$ in the soft band (0.5-2 keV) and 2.1$\times$10$^{-15}$ erg s$^{-1}$cm$^{-2}$ in the hard band (2-10 keV).
The full and hard band fluxes were extrapolated from net counts measured in 0.5-7 and 2-7 keV, respectively, assuming a power law with a slope of $\Gamma$=1.4 (not only for consistency with the work done in C-COSMOS, but also because this slope well represents a distribution of both obscured and unobscured AGN, being the X-ray background slope, see, e.g., Markevitch et al. 2003). We report in Table \ref{tab:num_src} the number of sources with DET\_ML$>$10.8 in at least one band, for each combination of bands.

\begin{table}[htbp]
\centering
\scalebox{1.}{
\begin{tabular}{cc}
\hline
\hline
Bands & Number\\
\hline
F+S+H & 1140\\
F+S & 536\\
F+H & 448\\
F & 121\\
S & 21\\
H & 7\\
\textbf{Total} & \textbf{2273}\\
\hline
\hline
\end{tabular}}\caption{Number of sources with DET\_ML$>$10.8 in at least one band, for each combination of X-ray bands (F: full; S: soft; H: hard). F+S+H: source with DET\_ML$>$6 in each band, and DET\_ML$>$10.8 in at least one band; F+S: source with DET\_ML$>$6 in full and soft bands, and DET\_ML$>$10.8 in at least one the two bands; F+H: source with DET\_ML$>$6 in full and hard bands, and DET\_ML$>$10.8 in at least one the two bands; F, S, H: sources with DET\_ML$>$10.8 only in full, soft, hard bands, respectively.}\label{tab:num_src}
\end{table}

We identify the X-ray sources using the same approach as C12, searching for counterparts in three different bands:
\begin{enumerate}
\item $i$ band ($\sim$7600 \AA), using the Subaru photometric catalog (Capak et al. 2007). Given that the Subaru catalog is saturated at magnitudes brighter than $i_{AB}$=20, we completed our $i$-band sample using information from the Canada-France-Hawaii Telescope (CFHT; McCracken et al. 2010) and from the SDSS catalog (see Section \ref{sec:srn_2-5} for further details of the positional match between the $i$-band CFHT and SDSS sources and the sources detected in $K$ or 3.6 $\mu$m band). In the analysis of the X-ray, optical and IR properties of the sample described in Section \ref{sec:opt_ir}, we used the Subaru magnitude; if the Subaru magnitude was not available, we used the CFHT magnitude, and we used the SDSS magnitude only for those sources with no Subaru or CFHT magnitude. Sources with only SDSS information are mainly very bright sources saturated in Subaru and CFHT catalogs. The final optical catalog contains about 870,000 sources at a signal-to-noise ratio (SNR) $>$5, covering a range in magnitude between $i\simeq$12 and $i\simeq$27. From now on we refer to this catalog as the ``optical catalog''. 
\item $K_S$ band (2.15 $\mu$m), using the UltraVISTA information from the Laigle et al. (submitted) catalog, not available at the time of C12, and the CFHT catalog. The UltraVISTA catalog has been obtained detecting and selecting objects using the ultra-deep chi-squared combination of $Y J H K_S$ and $z^{++}$ images. This catalog, although not $K$-selected, is sensitive to redder wavelengths than the Subaru $i$-band catalog, and it is therefore complementary to it. The catalog contains $\simeq$415,000 sources detected at SNR$>$5 to a $K_S$ magnitude limit of 26, and covers an area of $\simeq$2.0 deg$^2$, while the CFHT catalog contains $\simeq$320,000 sources detected at SNR$>$5 to a magnitude limit of 24.5, and covers an area of $\simeq$2.2 deg$^2$. In the analysis of the X-ray, optical and IR properties of the sample described in Section \ref{sec:opt_ir} we used the CFHT information only for sources with no secure UltraVISTA counterpart available. The \textit{Chandra} COSMOS Legacy survey area is not completely covered by the $K$-band catalog: 27 X-ray sources ($\simeq$1\%) are in fact outside the field of view of both the UltraVISTA and the CFHT surveys.
\item 3.6 $\mu$m, using the $Spitzer$ IRAC catalog from Sanders et al. (2007; hereafter we refer to this catalog as the Sanders catalog) and the SPLASH IRAC magnitude from the Laigle et al. (submitted) catalog (hereafter ``SPLASH catalog''). It is worth noticing that the SPLASH catalog, unlike the Sanders catalog, is not a 3.6 $\mu$m-selected catalog. The 3.6 $\mu$m SPLASH magnitude has been obtained performing aperture photometry at the position where the source has been detected in the combined $Y J H K_S$ and $z^{++}$ image. Nonetheless, we used the SPLASH information because it reaches more than 1.5 magnitudes deeper than the Sanders catalog, with a significantly smaller photometric error. The SPLASH catalog contains $\simeq$350,000 sources with SNR$>$5, with a magnitude limit of 26.0 (i.e., $\simeq$0.15 $\mu$Jy), and covers an area of $\simeq$2.4 deg$^2$: 22 \cha \leg sources lie outside the field of view of this catalog. The Sanders catalog contains instead $\simeq$330,000 sources at 3.6 $\mu$m to a magnitude limit of 24.5 (i.e., $\simeq$0.6 $\mu$Jy) at SNR$>$5 and covers the whole \cha \leg field. In the analysis of the X-ray, optical and IR properties of the sample described in Section \ref{sec:opt_ir}  we used the Sanders information only for sources with no secure SPLASH counterpart.
\end{enumerate}

In the final part of the identification process we also made use of the Advanced Camera for Surveys ($ACS$)/Hubble Space Telescope ($HST$) images of the COSMOS field (Scoville et al. 2007b; Koekemoer et al. 2007) to visually check our identifications, taking advantage of the ACS PSF, of the accuracy of the positions, and of the depth of the observations (I$_{F814W}\simeq$ 27.8 AB mag, 5$\sigma$ for an optimally extracted point source). The $ACS$/$HST$ survey covers only the central $\simeq$1.5 deg$^2$ of the COSMOS field, therefore only $\simeq$70\% of the \cha \leg sources were actually imaged with \textit{ACS}/\textit{HST}: for the remaining part, we used the $i$-band Subaru images.

We report in Table \ref{tab:mag_lim} the limiting magnitudes at SNR$>$5 for all the catalogs used in our identification process.

\begin{table}[h]
\centering
\begin{tabular}{cc}
\hline
\hline
Catalog & Mag$_{lim}$ (AB)\\
\hline
$i$ Subaru & 27.4\\
$i$ CFHT & 25.1\\
$i$ SDSS & 24.6\\
$K$ UltraVISTA & 26.0\\
$K$ CFHT & 24.0\\
3.6 $\mu$m SPLASH & 26.0\\
3.6 $\mu$m Sanders & 24.5\\
\hline
\hline
\end{tabular}\caption{Catalogs used to find Legacy counterparts and their magnitude limit at SNR$>$5.}\label{tab:mag_lim}
\end{table}

\section{X-ray source identification method}\label{sec:method}
\subsection{Method}
Following the procedure of Brusa et al. (2005), we used the likelihood ratio (LR) technique adopted in C12 and first developed by Sutherland \& Saunders (1992). This procedure was applied first to the XMM-COSMOS catalog (Brusa et al. 2007, hereafter B07; B10) with a percentage of ``reliable identifications'' greater than 80\%, and later on C-COSMOS with a percentage of ``reliable identifications'' of $\simeq$96\%. This technique takes into account, for each possible counterpart, the probability that it is a real or a spurious identification, using both the separation between the optical and the X-ray positions, and, as a prior, the information on the counterpart magnitude with respect to the overall magnitude distribution of sources in the field, thus making this method much more statistically accurate than one based on a positional match only.

The LR is defined as the ratio between the probability that an optical or infrared source is the correct identification and the corresponding probability for a background, unrelated object:
\begin{equation}
LR=\frac{q(m)f(r)}{n(m)}
\end{equation}

where $m$ is the magnitude and $r$ the positional offset from the X-ray source position of the optical or infrared candidate counterpart.

$n(m)$ is the density of background objects with magnitude $m$: we computed the distribution of the local background objects using the objects within a 5$^{\prime\prime}$--30$^{\prime\prime}$ annulus around each X-ray source. The 5$^{\prime\prime}$ inner radius was used in order to avoid the presence of true counterparts in the background distribution, while we chose a 30$^{\prime\prime}$ outer radius to avoid true counterparts of other X-ray sources. In the case of X-ray pairs the outer radius could contain the counterpart of a nearby X-ray source, but every annulus contains a number of background sources large enough ($\sim$80 sources in $i$-, $\sim$70 in $K$- UltraVISTA and $\sim$45 in the 3.6 $\mu$m band SPLASH catalog, respectively) to avoid significant effects of contamination.

$q(m)$ is the expected distribution function (normalized to 1) for the magnitude, $m$, of the real optical counterpart candidates. To compute $q(m)$ we first assumed an universal optical/infrared magnitude distribution for all X-ray sources, thus neglecting any influence of the X-ray flux on $q(m)$. Then we computed $q'(m)$ as the number of sources with magnitude $m$ within 1$^{\prime\prime}$ of the X-ray source, minus the expected number of background sources with magnitude $m$ in a 1$^{\prime\prime}$ circle. The 1$^{\prime\prime}$ radius maximizes the statistical significance of the overdensity around X-ray sources: a smaller radius would give a higher Poissonian noise, while a larger radius would increase the number of background sources. 
Finally, we normalized $q'(m)$ in order to have $q(m)$= $const$ $\times$ $q'(m)$ such that $\int_{-\infty}^{+\infty} q(m)\, dm$ = 1. The normalization value $const$ is here assumed 0.92, slightly larger than in C12, where it was $const$=0.9. This normalization choice is the best trade-off between completeness and reliability, i.e., it allows us to find a larger number of counterparts without significantly increasing the number of expected spurious detections.

Finally, $f(r)$ is the probability distribution function of the positional errors, assumed to be a two-dimensional Gaussian, with $\sigma$=$\sqrt{\sigma_X^2+\sigma_{opt}^2}$. $\sigma_X$ is the X-ray positional uncertainty, computed as described in Paper I, while $\sigma_{opt}$ is the positional uncertainty in the optical/IR band. We adopted the same optical positional uncertainties of C12, i.e., 0.2$^{\prime\prime}$ for the $K$ band (McCracken et al. 2010), 0.3$^{\prime\prime}$ for the $i$ band (Capak et al. 2007) and 0.5$^{\prime\prime}$ for the 3.6 $\mu$m band (Sanders et al. 2007).

\subsection{Threshold Choices}
A fundamental step in the optical counterparts identification is the choice of the best likelihood threshold value ($L_{th}$) for LR, in order to make a distinction between real and spurious identifications. $L_{th}$ should not be too high, otherwise we would miss too many real identifications and consequently reduce the sample completeness, but $L_{th}$ has also to be high enough to keep the number of spurious identifications low and the reliability of the identification high.

Reliability describes the possibility of having multiple candidate counterparts for the same X-ray source. For a given optical object $j$, the reliability $R_j$ of being the correct counterpart is
\begin{equation}
R_j=\frac{(LR)_j}{\sum_i{(LR)_i+(1-Q)}},
\end{equation}

where the sum is over the set of all optical candidate counterparts  and $Q=\int_{m} q(m)\, dm$ is normalized in order to be equal to the ratio between the number of X-ray sources identified in the given optical/infrared band and the total number of sources in the X-ray sample. The reliability $R_k$ for each X-ray source is the sum of the reliabilities $R_{j}$ of all the possible counterparts of the $k$-th X-ray source and it is by definition equal to 1. The reliability parameter ($R$) for the whole sample, instead, is defined as the ratio between the sum of all the reliabilities of the candidate counterparts and the total number of sources with LR$>$$L_{th}$, i.e. R=$N_{ID}$/$N_{LR>L_{th}}$. 

The completeness parameter ($C$) of the total sample is defined as the ratio between the sum of the reliability of all the sources identified as possible counterparts and the total number of X-ray sources ($C$=$N_{ID}$/$N_{X}$).

\begin{figure}[H]
\centering
\includegraphics[width=0.5\textwidth]{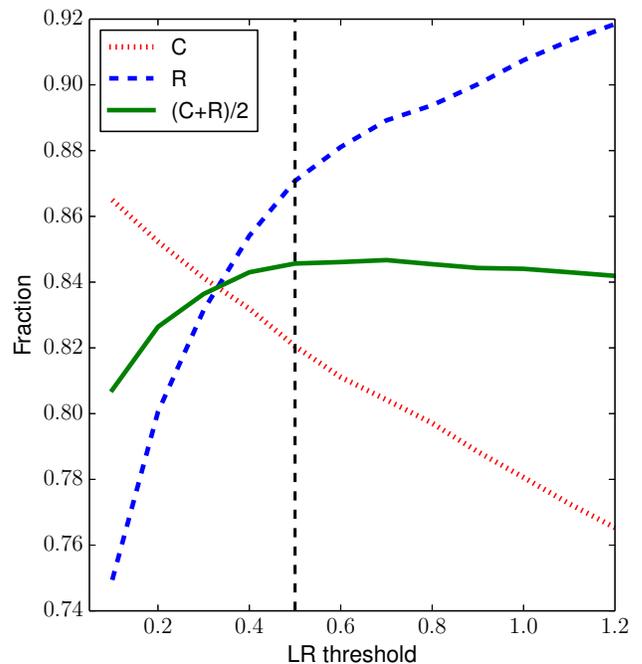}
\caption{Completeness (C, red dotted line), reliability (R, blue dashed line) and (C+R)/2 (green solid line) at given values of L$_{th}$ matching the optical catalog  with new Legacy sources. The dashed black line shows the selected threshold in this band, L$_{th}$=0.5}\label{fig:cr_092_leg_i_3band}
\end{figure}

In C12 and in B07, $L_{th}$ was defined as the likelihood ratio where the quantity ($C+R$)/2 is maximized. In the \cha \leg survey ($C$+$R$)/2 is almost flat at L$_{th}$$\geq$0.5, as can be seen in Figure \ref{fig:cr_092_leg_i_3band}, where we plot $C$, $R$ and ($C$+$R$)/2 versus L$_{th}$ for the optical catalog, so we select a $L_{th}$ value of 0.5 for both $i$ and $K$-bands. Given the lower spatial resolution of the 3.6 $\mu$m data, we chose a slightly higher threshold $L_{th}$=0.7 in this band, to reduce the number of spurious identifications. The corresponding sample completeness and reliability for the catalogs in the three bands are shown in Table \ref{tab:c_r}: as a general trend, both C and R grow moving from optical to infrared, due to the stronger relation of $K$ or 3.6 $\mu$m magnitudes with the X-ray flux (Mainieri et al. 2002, Brusa et al. 2005).

As a final remark, it is worth noticing that the values of $C$ and $R$ we obtained for the new \textit{Chandra} COSMOS Legacy dataset are all in good agreement with those obtained for C-COSMOS ($C$=0.85 and $R$=0.88 for $i$, $C$=0.90 and $R$=0.92 for $K$, and $C$=0.96 and $R$=0.96 for 3.6 $\mu$m), and are higher than those of XMM-COSMOS because of the better \textit{Chandra} positional accuracy (angular resolution of $\simeq$0.5$^{\prime\prime}$ and $\simeq$6$^{\prime\prime}$ for \cha and \xmm full width half maximum, FWHM, respectively).

\begin{table}[H]
\centering
\scalebox{1.}{
\begin{tabular}{cccc}
\hline
\hline
Band & C & R & LR$_{th}$\\
\hline
$i$ & 0.82 & 0.87& 0.5\\
$K$ & 0.86 & 0.93 & 0.5 \\
3.6 $\mu$m & 0.92 & 0.97 & 0.7\\
\hline
\hline
\end{tabular}}\caption{Completeness (C) and Reliability (R) for each optical/IR band.}\label{tab:c_r}
\end{table}

\section{X-ray source identification results}\label{sec:results}
In this Section we show the procedure adopted to define the final counterparts after performing the likelihood ratio analysis. As in C12 and in XMM-COSMOS (B07, B10), the X-ray sources have been divided into four classes, based on their counterparts associations:
\begin{enumerate}
\item Secure. Sources with only one counterpart with LR$>$LR$_{th}$. The vast majority of counterparts belongs to this class. 2214 of the 2273 new \cha \leg sources ($\simeq$97\%) have been classified secure after the whole identification procedure (see Table \ref{tab:final_opt_match}).
\item Ambiguous. Sources with more than one counterpart  above the threshold. 24 of the 2273 new \cha \leg sources have been classified as ambiguous after the whole identification procedure.
\item Subthreshold. Sources with one or more possible counterparts with LR$<$LR$_{th}$ within 5$^{\prime\prime}$ from the X-ray centroid. 4 of the 2273 new \cha \leg sources have been classified as subthreshold after the whole identification procedure.
\item Unidentified. Sources with no counterpart, even below the threshold, within 5$^{\prime\prime}$ from the X-ray centroid. 31 of the 2273 new \cha \leg sources have been classified as unidentified after the whole identification procedure.
\end{enumerate}

A few examples of objects belonging to these classes are shown in Figure 3 of C12.

\subsection{Identification rates}
First of all, we run the LR technique with the $K$-band catalogs, using both the UltraVISTA and the CFHT catalogs we described in Section \ref{sec:datasets}: the positional error for the $K$-band sources has been fixed to 0.2$^{\prime\prime}$, as in C12.
We first matched our sources with those in the UltraVISTA area, assuming $L_{th}$=0.5, and we obtained 1690 counterparts with $LR>L_{th}$, while another 117 sources have a counterpart with LR$<$L$_{th}$. 583 \cha \leg sources have therefore no secure UltraVISTA counterpart (117 sources with a counterpart with LR$<$L$_{th}$ and 466 with no UltraVISTA counterpart). In the CFHT catalog, 379 of these 583 sources have at least one counterpart with $LR>L_{th}$: as a final summary, 2069 sources (92.2\% of the X-ray sample inside the composite $K$-band field of view) have at least one counterpart with $LR>L_{th}$ in the $K$-band.

We then run the LR technique with the $i$-band Subaru catalog we described in Section \ref{sec:datasets}. The adopted positional error for the $i$-band sources is 0.3$^{\prime\prime}$, as in C12. 
At a $L_{th}$ value of 0.5, there are 1594 Legacy sources (70.1\%) with secure or ambiguous Subaru $i$-band counterpart  with SNR$>$5 and $LR>L_{th}$, while another 69 sources (3.0\% of the whole sample) have one or more counterparts with $LR<L_{th}$.

Finally, we matched our X-ray catalog with the 3.6 $\mu$m catalog: the positional error for the 3.6 $\mu$m sources has been fixed to 0.5$^{\prime\prime}$, as in C12. We first matched the X-ray catalog with the SPLASH catalog:  at a $L_{th}$ value of 0.7, there are 2046 Legacy sources with at least one SPLASH counterpart  with SNR$>$5  and $LR>L_{th}$ (91.1\% of 2246 X-ray sources inside the SPLASH field of view), while another 41 sources (1.8\%) have one or more counterparts with LR$<$L$_{th}$.  227 \cha \leg sources have therefore no secure SPLASH counterpart (41 sources with a counterpart with LR$<$L$_{th}$ and 186 with no UltraVISTA counterpart). We then matched these 227 sources, with the Sanders catalog, and we found another 125 sources with  $LR>L_{th}$. Therefore, combining the two 3.6 $\mu$m catalogs 2171 sources (95.5\% of the whole sample) have at least one counterpart with $LR>L_{th}$. 

The identification rates in all bands are in very good agreement with those reported in C12.

\subsection{Counterparts with 2$<$SNR$<$5}\label{sec:srn_2-5}
In order to complete our identification of optical counterparts, we looked for $i$ and $K$-band counterparts with 2$<$SNR$<$5; we did not perform this analysis in the 3.6 $\mu$m band, due to its lower spatial resolution. There are 157 X-ray sources with no counterpart with SNR$>$5 in $i$-band but with at least one counterpart with 2$<$SNR$<$5 in $i$-band. Of these sources, 148 have at least one counterpart with $LR>L_{th}$, while the other 9 have LR$<$L$_{th}$. There are also 18 X-ray sources with no counterpart with SNR$>$5 in $K$-band but at least one counterpart with 2$<$SNR$<$5 and $LR>L_{th}$ in the composite UltraVISTA/CFHT $K$-band, and one source with one counterpart with 2$<$SNR$<$5 and LR$<$L$_{th}$ in the composite UltraVISTA/CFHT $K$-band.

To complete our $i$-band catalog, especially at $i_{AB}$$<$20, where the Subaru catalog is saturated, we also matched our $K$ and 3.6 $\mu$m secure counterparts with the CFHT and SDSS $i$-band catalogs, with maximum separation $d_{ik}$=1$^{\prime\prime}$: we found $i$-band magnitude for 301 X-ray sources (13.2\% of the whole sample).

We report in Table \ref{tab:first_opt_match} the number of counterparts in the i, K and 3.6 $\mu$m bands, first using only sources with SNR$>$5, then introducing also sources with 2$<$SNR$<$5. As can be seen, the fraction of sources with a secure counterpart is excellent in every band (79.8\% in the $i$-band, 85.1\% in the $K$-band and 90.1\% in the 3.6 $\mu$m band), but the number of ambiguous sources, i.e. of sources with more than one possible counterpart in an optical or IR band, is significant, especially in the $i$ and $K$-bands, where $\simeq$9\% and $\simeq$7\% of the \cha \leg sources are ambiguous. In the next section, we describe the approach chosen to significantly reduce the number of ambiguous counterparts.

\subsection{Solving the cases of ambiguous sources}\label{sec:amb_solve}
As previously explained, $\simeq$8\% of X-ray sources have been flagged as ``ambiguous'' in both $i$ and $K$-band. We developed the following procedure to choose the correct counterpart: the main assumption is to use secure counterparts in one band to solve ambiguities in the other one. We started by matching $i$ and $K$-band counterparts, and then we introduced those in the 3.6 $\mu$m. For each source we run a four different checks: if one was not satisfied, we moved to the following one.

\begin{enumerate}
\item Counterparts in $i$ and $K$-band have R$>$0.9 in both bands. We kept these counterparts as the good ones and we rejected any other counterpart of the same X-ray source. The largest part of ambiguities ($\simeq$50\%) is solved in this first step.
\item There is a counterpart with R$>$0.9 in one band and the distance between this counterpart and only one counterpart in the other band is $d_{ik}<$1$^{\prime\prime}$. We kept these counterparts as the good ones and we reject any other counterpart of the X-ray source. Other $\simeq$25\% of ambiguities is solved in this step. 
\item The two counterparts with largest R have $d_{ik}<$1$^{\prime\prime}$. We kept these counterparts as the good ones and we rejected any other counterpart of the same X-ray source. After this step, less than 15\% of the original ambiguous identifications are still ambiguous.
\item There is a secure 3.6 $\mu$m counterpart within 1$^{\prime\prime}$ from the X-ray source and one of the counterparts in $i$ or $K$-band have distance from the 3.6 $\mu$m counterpart $<$1$^{\prime\prime}$.
\end{enumerate}

The number of 3.6 $\mu$m ambiguous identifications is lower than in the $i$ and $K$-band ones, because of the \textit{Spitzer} lower spatial resolution. For this reason, to solve ambiguities in the IRAC band we adopted a simplified procedure, where we first kept, if present, the sources with R$>$0.9 and rejected the other candidates. Then, for the smaller fraction of sources still ambiguous ($\simeq$15 sources), we looked for a secure counterpart in the optical or $K$-band within 1$^{\prime\prime}$. With this procedure, no counterpart in the 3.6$\mu$m band is flagged as ambiguous. 

During the analysis of ambiguous sources, we did not use deblending techniques.

\subsection{Final results for optical counterparts}
We finally performed a complete visual check of all the X-ray sources and their counterparts: we found a further group of visually good counterparts ($\simeq$2\% of all the secure counterparts in the optical catalog, and $\simeq$1\% of all the secure counterparts in the $K$-band catalog), which were not previously found mainly because they had SNR$<$2. All these new counterparts have separation from the X-ray centroid smaller than 1$^{\prime\prime}$ and already have a counterpart detected with the LR ratio technique in at least one of the other two optical/IR bands. 

We report in Table \ref{tab:final_opt_match} the final number of counterparts in the i, K and 3.6 $\mu$m bands, after the resolution of ambiguous counterparts and the visual inspection. 2214 sources (97.4\%) have now a secure counterpart, i.e., one counterpart above LR$>$L$_{th}$ with all the possible others above threshold rejected after our procedure and visual inspection: this result is comparable with the one obtained in CDF-S (96.8\%, Xue et al. 2011) and better than the one in Stripe 82 ($\simeq$80\% in the optical SDSS band, $\simeq$59\% in the UKIDSS near-IR band and $\simeq$65\% in the WISE 3.6 $\mu$m band, LaMassa private communication). Other 24 sources (1.1\%) have been instead classified as ambiguous, and only four sources are classified as subthreshold. Finally, 31 sources (1.4\%) have no counterpart in any of the optical or infrared bands. These sources are candidate obscured or high-z AGN, or both; however, it is also worth noticing that a fraction of 0.3\% of \cha \leg sources (i.e. $\simeq$12 in full, 9 in soft and 8 in hard band, assuming a threshold of 7 net counts) is expected to be spurious at the likelihood threshold used in the X-ray catalog (Paper I).

We also point out that the fraction of counterparts we found is consistent with the one obtained by Hsu et al. (2014, $\simeq$96\%) using a slightly different matching method, based on Bayesian statistics, which also takes into account both the magnitude and the source position, as the LR ratio technique, and in addition works simultaneously on multiple bands. We decided not to use this technique for consistency with the C-COSMOS analysis and also because the Hsu et al. (2014) method, although used on the CDF-S, becomes significantly more effective than the one we used only on very large area surveys, with millions of potential counterparts, a significant fraction of which with non-negligible positional error and without homogenous multiwavelength coverage.

\begin{table*}[!h]
\centering
\scalebox{1.}{
\begin{tabular}{cccccccccccc}
\hline
\hline
Class & $i_{p}$ & $i_{p}$ & $i_{other}$ & $i_{whole}$ & f$_{i,whole}$ & $K$ & $K$ & f$_{K}$  & 3.6 $\mu$m & 3.6 $\mu$m & f$_{3.6}$\\ 
& SNR$>$5 & SNR$>$2 & & SNR$>$2 &  SNR$>$2 & SNR$>$5 & SNR$>$2 & SNR$>$2 & SNR$>$5 & SNR$>$2& SNR$>$2\\ 
\hline

Secure & 1465 & 1581 & 232 &1813 & 79.8\% &  1923 & 1935  & 85.1\% & 2049 & 2049 & 90.1\%\\ 
Ambiguous & 129 & 161 & 40 & 201 & 8.8\% &  148 & 154 & 6.8\% & 125 &  125 & 5.5\%\\ 
Subthreshold & 69 & 78 & 29 & 107 & 4.7\%\ & 53 & 54 & 2.4\% &  37 & 37 & 1.6\%\\  
Unidentified & 610 & 453 & -- & 152 & 6.7\% & 149 & 130 & 5.7\% & 62 & 62 & 2.7\%\\
\hline
\hline
\end{tabular}}\caption{Number of X-ray sources identified in each band and in total, for counterparts with SNR$>$5 and adding counterparts with 2$<$SNR$<$5, and fraction $f$ of sources with respect to the whole survey, after the contribution of sources with 2$<$SNR$<$5 has been taken in account. $i_{p}$ identifies sources with Subaru $i$-band magnitude, $i_{other}$ identifies sources with CFHT or SDSS $i$-band magnitude and $i_{whole}$ summarizes all sources with $i$-band magnitude.}\label{tab:first_opt_match}
\end{table*} 

\begin{table*}[!h]
\centering
\scalebox{1.}{
\begin{tabular}{ccccccccc}
\hline
\hline
Class & $i$ & f$_{i}$ & $K$ & f$_{K}$ & 3.6 $\mu$m & f$_{K}$ & Total & f$_{total}$\\
\hline
Secure & 2100 & 92.4\% &  2119  & 93.2\% & 2171 & 95.6\% & 2214 & 97.4\% \\  
Ambiguous & 17 & 0.7\% &  9 & 0.4\% & 0 & 0\% & 24 & 1.1\%\\  
Subthreshold & 92 & 4.0\%\ & 28 & 1.3\% & 36 & 1.6\% & 4 & 0.1\% \\  
Unidentified & 64 & 2.8\% & 117 & 5.1\% & 66 & 2.7\% & 31 & 1.4\%  \\  
\hline
\hline
\end{tabular}}\caption{Final number of X-ray sources identified in each band and in total. ``Total'' is the number of sources with an identification in one or more bands.}\label{tab:final_opt_match}
\end{table*}

\subsection{Sources in C-COSMOS with updated optical counterpart}
676 of the 1743 C-COSMOS sources have been observed again during the \cha \leg observations, thus having now double \cha exposure and therefore improved positional accuracy $err_{pos}$, given that $err_{pos}\propto C_{S}^{-0.5}$, where $C_{S}$ are the source net counts (see Paper I for further details). We performed the same LR technique we used on the 2273 new \cha \leg sources to check if the potential slight change in the X-ray position of the source due to the larger exposure and/or the use of different catalogs of optical/IR counterparts with respect to C12 implied the identification of a different (or new) optical/IR counterpart. We found that 9 (1.3\%) sources have a different optical/IR counterpart, while 6 (1\%) sources that had no optical counterpart in C-COSMOS now have a secure one. For all these sources, a new photometric redshift (see Section \ref{sec:photoz}) has also been computed. 

We also run the LR ratio identification procedure on the whole C-COSMOS sample with the new UltraVISTA and SPLASH information, that were not available at the time of C12. We found 52 sources (3\% of the whole C-COSMOS sample) with no CFHT $K$-band information but with a secure UltraVISTA counterpart. We also found 49 sources (2.8\% of the whole C-COSMOS sample) with no 3.6 $\mu$m IRAC information from the Sanders catalog, but with a secure SPLASH counterpart.

We report this updated information, together with the newly available redshifts, in the new catalog of optical counterparts of the whole \cha \leg survey.

\subsection{Sources in XMM-COSMOS with updated optical counterpart}
866 new \cha \leg sources have a counterpart in the XMM-COSMOS catalog (Cappelluti et al. 2009) within 10$^{\prime\prime}$. 104 of these sources have a different optical counterpart than in the XMM-COSMOS optical catalog (B10), mainly because of the better angular resolution of \cha compared to XMM-\textit{Newton}, but also because we used different optical catalogs; also, a significant fraction of these sources was flagged as ambiguous in B10. It is also worth noticing that 36 of these \cha sources are actually part of pairs of counterparts of the same \xmm source (once again because of the better \cha angular resolution).

We report the XMM-COSMOS identification number of all \cha \leg sources in the new catalog available with this Paper, in the COSMOS repository and online.

\section{Positional offset and optical properties by identification class}\label{sec:id_prop} 

\begin{figure*}[!h]
  \begin{minipage}[b]{.5\linewidth}
    \centering
    \includegraphics[width=1.03\textwidth]{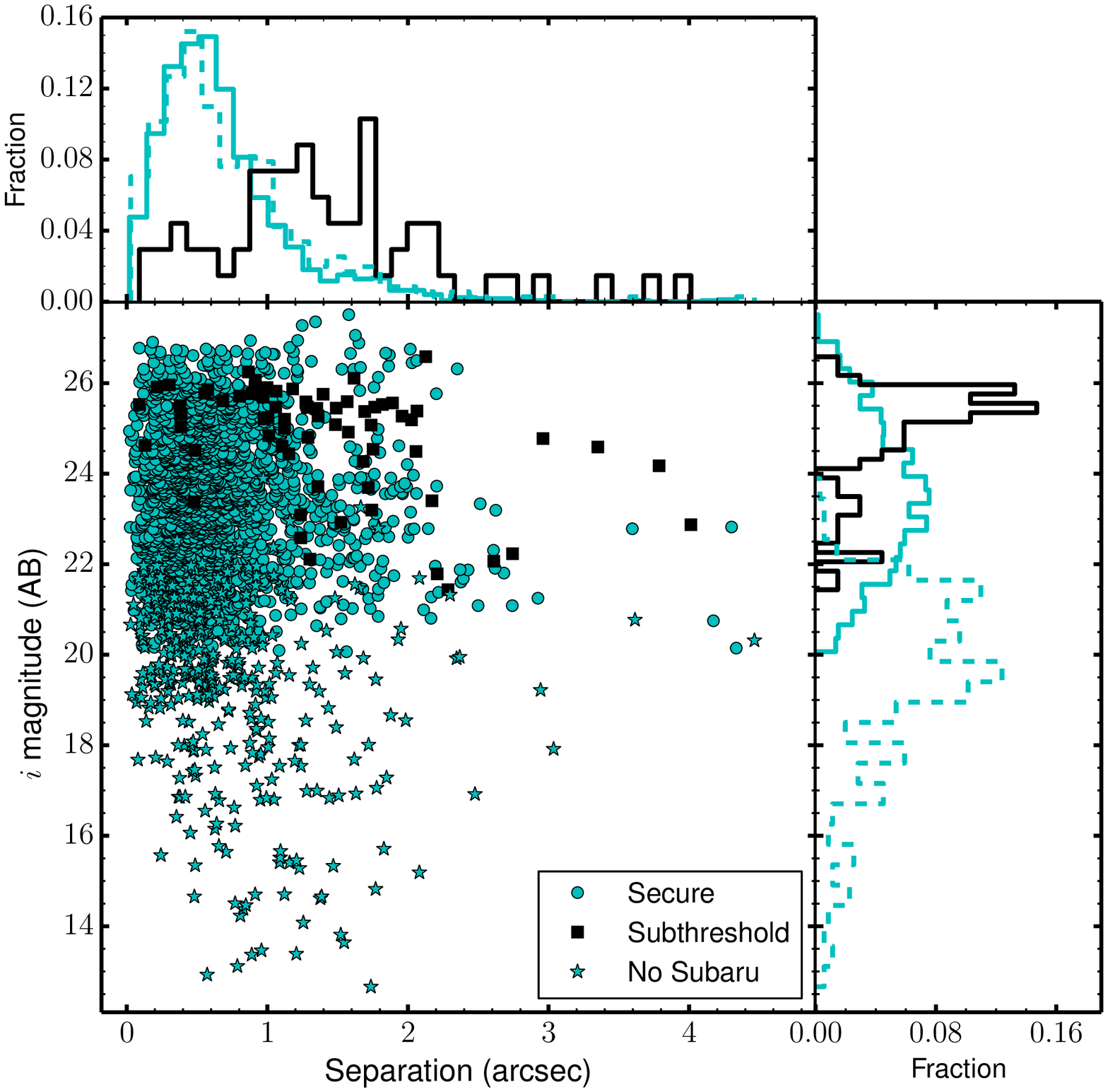}
  \end{minipage}
  \hfill
  \begin{minipage}[b]{.5\linewidth}
    \centering
    \includegraphics[width=1.01\textwidth]{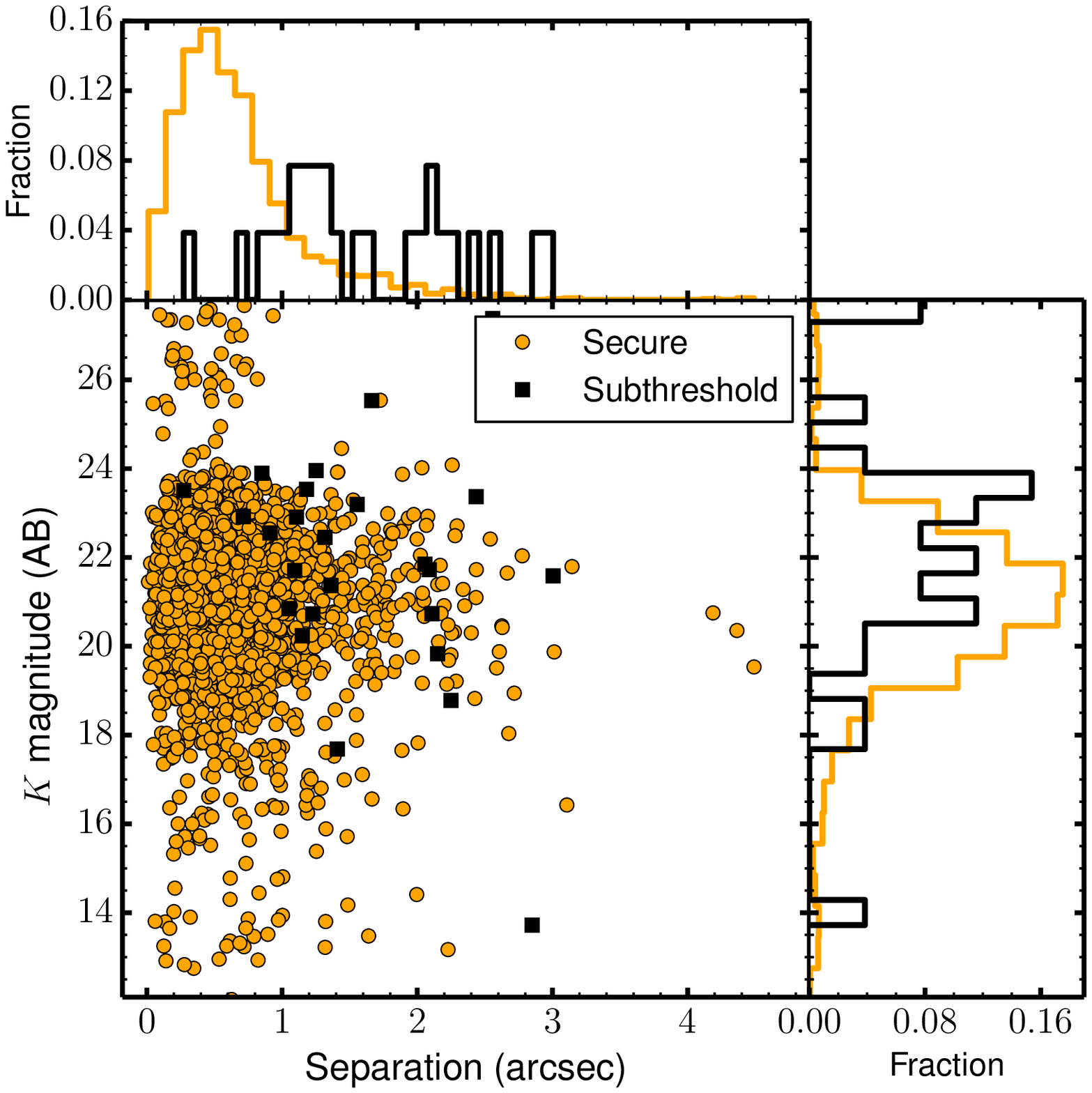}
  \end{minipage}
    \begin{minipage}[b]{.99\linewidth}
    \centering
    \includegraphics[width=0.5\textwidth]{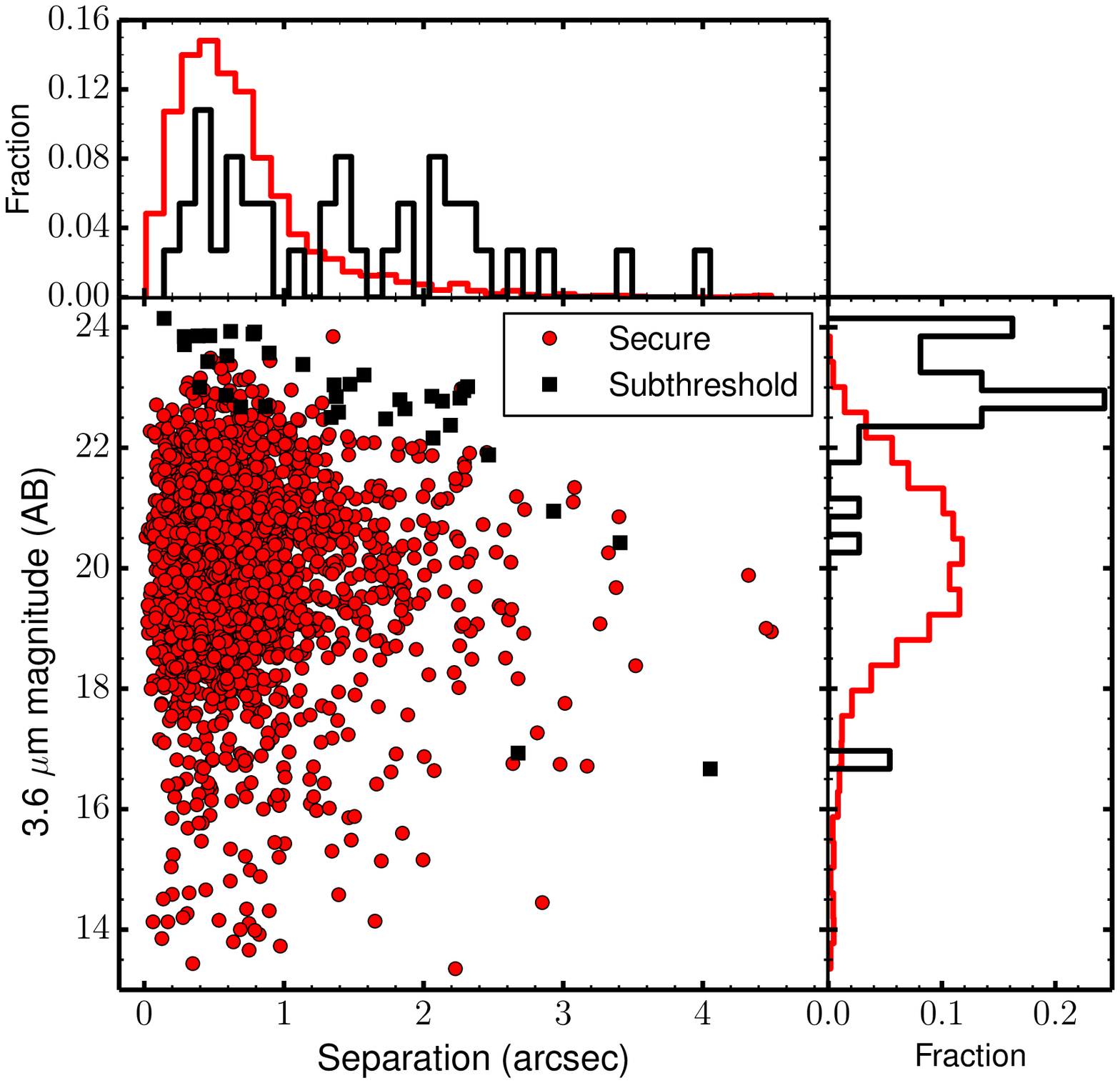}
  \end{minipage}
\caption{Separation between X-ray and optical (IR) positions for $i$-band (top left), $K$-band (top right), and 3.6 um bands (bottom center). Secure counterparts are shown in cyan ($i$-band), orange ($K$-band) and red (3.6 $\mu$m band), while sub-threshold counterparts are shown in black. Sources with $i$-band magnitude from CFHT or SDSS are plotted as cyan stars. Histograms of separation and magnitude are shown in each of the three plots. Histogram of separation and magnitude for $i$-band sources with CFHT or SDSS information are showed with a dashed line in the $i$-band diagram.}
\label{fig:sep_vs_mag_x_secure}
\end{figure*}

We present in Figure \ref{fig:sep_vs_mag_x_secure} the X-ray to optical/IR separation and the optical/IR magnitude,  in $i$ (top left panel, cyan), $K$ (top right panel, orange) and 3.6 $\mu$m band (bottom center panel, red). 
More than 90\% of the secure counterparts have a distance from the X-ray source smaller than 1$^{\prime\prime}$: the mean (median) value of the distance from the X-ray source is 0.70$\pm$0.50$^{\prime\prime}$ ( 0.59$^{\prime\prime}$) for the $i$-band counterparts, 0.67$\pm$0.48$^{\prime\prime}$ (0.56$^{\prime\prime}$) for the $K$-band counterparts and 0.69$\pm$0.49$^{\prime\prime}$ (0.58$^{\prime\prime}$) for the 3.6 $\mu$m counterparts, a result in agreement with the one obtained during the astrometric correction process of the X-ray observations described in Paper I. The distribution is instead wider for subthreshold sources, where the mean (median) distance from the X-ray source is  1.24$\pm$0.70$^{\prime\prime}$ (1.17$^{\prime\prime}$) in $i$-band, 1.64$\pm$0.67$^{\prime\prime}$ (1.56$^{\prime\prime}$) in $K$-band and 1.60$\pm$0.96$^{\prime\prime}$ (1.57$^{\prime\prime}$) in 3.6 $\mu$m band. Moreover, subthreshold counterparts are on average 1.5-2 magnitudes fainter than the secure counterparts (see Table \ref{tab:mag_prop} for a summary). Both the fainter magnitudes and the larger X-ray to optical/IR separations are consistent with the subthreshold counterparts being less reliable than the secure ones (see also B10).

\begin{table}[H]
\centering
\scalebox{0.95}{
\begin{tabular}{ccccccccc}
\hline
\hline
Band &\multicolumn{2}{c}{Secure} & \multicolumn{2}{c}{Subthreshold}\\
& mean & median & mean & median\\
\hline
$i$ & 22.8 & 23.0 & 24.5 & 25.2\\ 
$K$ & 20.9 & 21.0 & 22.1 & 22.8\\  
3.6 $\mu$m IRAC & 20.4 & 20.5 & 22.4 & 22.8\\  
\hline
\hline
\end{tabular}}\caption{Mean and median magnitude values for secure and subthreshold counterparts in each of the three bands used in our analysis.}\label{tab:mag_prop}
\end{table} 

We also analyzed the distribution of the distance between optical and infrared counterparts for the same X-ray source: for secure counterparts (in both bands), the mean (median) distance between $i$ and $K$ counterparts is 0.17$\pm$0.38$^{\prime\prime}$ (0.07$^{\prime\prime}$) and that between $i$ and 3.6 $\mu$m counterparts is 0.17$\pm$0.41$^{\prime\prime}$ (0.07$^{\prime\prime}$). We do not report the distance between secure $K$ and 3.6 $\mu$m counterparts because the vast majority of them come from the same catalog (Laigle et al. submitted), which contains both the UltraVISTA and the SPLASH magnitude information, and have therefore the same right ascension and declination. This small value in the separation between optical and $K$/IR counterparts is consistent with the fact that secure counterparts in different bands are actually the same source.

We studied the distribution of the difference between X-optical distances of the closest and the second closest possible counterpart of ``ambiguous'' identifications, and the distribution of the difference $\|$Mag$_2$-Mag$_1$$\|$, where Mag$_1$ and Mag$_2$ are the magnitudes of the ``ambiguous'' identifications: here we define as ``ambiguous'' only the sources with no secure counterpart after running the procedure described in Section \ref{sec:amb_solve}. As expected, for more than 75\% of the ambiguous counterparts the difference between the distances of the two candidate counterparts from the X-ray source is smaller than 1$^{\prime\prime}$, i.e. comparable with \cha resolution ($\simeq$0.5$^{\prime\prime}$). Similarly, for more than 70\% of the ambiguous identifications the difference in magnitude between the two candidate counterparts is $<$1 mag, therefore not allowing to select one of the sources as a secure counterpart.

\section{Spectroscopic and photometric redshift distribution}\label{sec:z_prop}
\subsection{Spectroscopic redshifts}
We cross-correlated our optical counterparts with the master spectroscopic catalog available within the COSMOS collaboration (Salvato et al. in prep.), which contains $\simeq$80,000 spectroscopic redshifts.. The catalog includes redshifts from SDSS (DR12),  VIMOS (zCOSMOS: Lilly et al. 2007, Lilly et al. 2009; VUDS: Le Fevre  et al. 2015), MOSFIRE (Scoville et al. 2015 in prep; MOSDEF: Kriek et al. 2015),  several years of DEIMOS observations from multiple observing programs (e.g. Kartaltepe et al. 2010, Hasinger et al. 2015 in prep.), IMACS (Trump et al. 2007, 2009a), Gemini-S (Balogh et al. in prep.), FORS2 (Comparat et al. 2015), FMOS (Silverman et al. 2015), PRIMUS (Coil et al. 2011) and HECTOSPEC (Damjanov et al. 2015), plus a negligible number of sources provided by other smaller contributions. 

The redshift confidence from the various contributors has been translated into the classification as defined in zCOSMOS: 730 of the 2273 \cha \leg sources have a reliable spectroscopic redshift, i.e. with confidence $\geq$3 (spectroscopic accuracy $>$99.5\%, estimated using those objects observed more than once, and verifying if their redshift were in agreement). However, our sample contains also 211 sources with a less reliable spectroscopic redshift (spectroscopic accuracy $<$99.5\%) but with the photometric redshift $z_{phot}$ specifically provided for this catalog (see Section \ref{sec:photoz}) such that $\Delta z=\frac{|z_{spec}-z_{phot}|}{1+z_{spec}}<0.1$. For these sources, we adopted as final value the spectroscopic redshift one. In summary, we provide a spectroscopic redshift for 941 sources ($\simeq$41\% of the sample).

\subsection{Photometric redshifts}\label{sec:photoz}
\begin{figure*}[!h]
  \begin{minipage}[b]{.5\linewidth}
    \centering
    \includegraphics[width=1.03\textwidth]{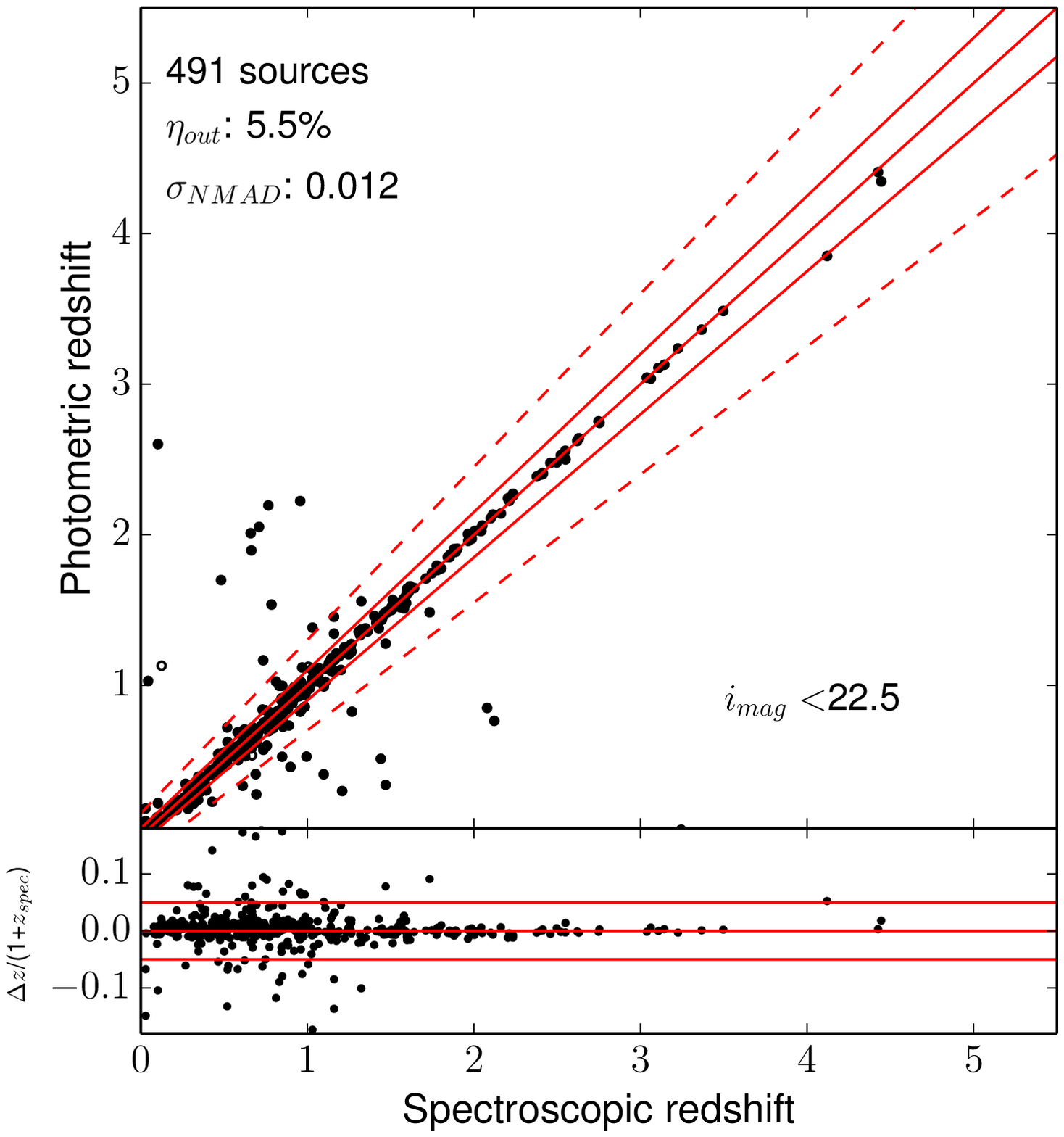}
    \label{subfig-1:i_band}
  \end{minipage}
  \hfill
  \begin{minipage}[b]{.5\linewidth}
    \centering
    \includegraphics[width=1.01\textwidth]{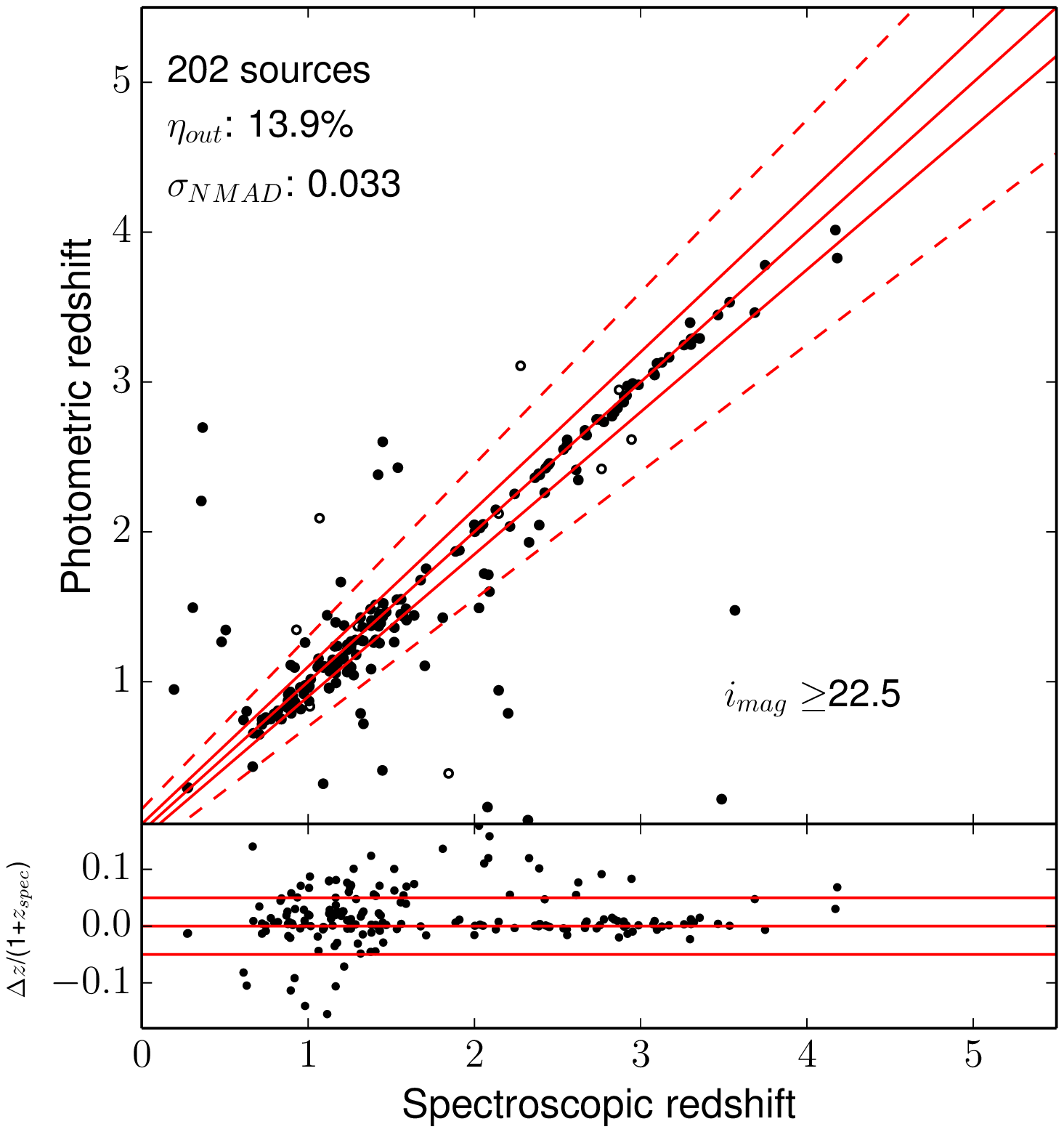}
    \label{subfig-2:k_band}
  \end{minipage}
\caption{Photometric redshifts compared to the secure spectroscopic redshifts, for sources brighter (left) and fainter (right) than $i_{AB}$=22.5. Open circles represent sources for which there is at least a second significant peak in the redshift probability distribution. Red solid lines correspond to zphot = zspec and zphot = zspec $\pm$0.05$\times$(1+zspec), respectively. The dotted lines limit the locus where zphot = zspec $\pm$0.15$\times$(1 + zspec). Photo-z computed for the fainter sources are significantly worse in terms of both dispersion and fraction of outliers.}\label{fig:spec_vs_phot}
\end{figure*}

For 1234 sources, we can provide only photometric redshifts. Photometric redshifts have been produced following the same procedure described in detail in Salvato et al. (2011), without any further training sample. Depending on the X-ray flux of the sources and on the  morphological and photometric (e.g variability) properties of the counterpart, specific priors and libraries of templates (including galaxies, AGN/galaxy hybrids, AGN and QSOs)  have been adopted, and the best fit has been found through a $\chi^2$ minimization, using the publicly available code LePhare (Arnouts et al. 1999, Ilbert et al. 2006). Using the secure spectroscopic subsample as reference (i.e. only those sources with spectroscopic accuracy $>$99.5\%), we found an accuracy of $\sigma_{\Delta z/(1+z_{spec})}$=0.03, with a fraction of outliers, i.e., sources with $\Delta z/(1+z_{spec})>$0.15, $<$8\% (55 of 699), consistent with what was already found for C-COSMOS (78 outliers out of 1020 secure spec-z), using a different spectroscopic sample. Breaking down the sample (see Figure \ref{fig:spec_vs_phot}), for the 491 sources that are brighter than $i_{AB}$=22.5 the accuracy, estimated using the normalized median absolute deviation $\sigma_{NMAD}$=1.48$\times$median($\|$$z_{spec}$-$z_{phot}$$\|$/(1+$z_{spec}$)), is $\sigma_{NMAD}$=0.012 with 5.5\% of outliers. For the fainter sample of 202 sources, where the number of the available photometric bands  decreases and the photometric errors increase, the accuracy decreases by a factor of $\sim$ 3 ($\sigma_{NMAD}$=0.033) and the number of outliers increases by the same factor (13.9\%). The whole sample has $\sigma_{NMAD}$=0.014 with 7.9\% of outliers.

The photo-z computation provides for each source a probability distribution function (Pdz), which gives the probability of a source to be at a given redshift bin: the nominal photo-z value is actually the maximum of this Pdz. The integrated area of the Pdz on all redshift bins is by definition equal to 1. At all redshifts, the agreement between the distribution of the nominal values of the photometric redshifts and the average distribution of the Pdz is very good. However, using the Pdz instead of just the photo-z nominal value allows to perform a much more statistically thorough analysis (Georgakakis et al. 2014). \cha \leg Pdz-s are already being used in the space density computation at $z>$3 (Marchesi et al. to be submitted) and in the AGN clustering estimation at high redshift (Allevato et al. in prep.).

\subsection{Redshift Summary}
From this point of the Paper, we will talk about the whole \cha \leg survey, i.e., of both the new dataset described so far, together with the old C-COSMOS sources.

The total number of new \cha \leg sources with a redshift, either spectroscopic or photometric, is 2182, i.e. 96\% of the entire sample. In C-COSMOS, 1695 of the 1743 X-ray sources have a redshift (i.e. 97.3\%), 1211 of which have a reliable spectroscopic redshift, either secure (1032) or in agreement with the photo-z (179). With respect to the C12 catalog, we added new reliable spectroscopic redshift information to 296 C-COSMOS sources. 

Summarizing, 3877 of the 4016 X-ray sources in the whole  \cha \leg field have a redshift, i.e. $\simeq$96.5\% of the whole sample. We have a reliable spectroscopic redshift for 2151 of these sources (53.6\% of the whole sample). As a comparison, $\simeq$91\% of the 740 sources in CDF-S (Xue et al. 2011) have either a spectroscopic or photometric redshift, and $\simeq$46\% have a reliable spectroscopic redshift, while $\simeq$30\% of the sources in Stripe 82 (LaMassa et al. 2013 and submitted) have a reliable spectroscopic redshift. 

\begin{figure}[H]
\centering
\includegraphics[width=0.5\textwidth]{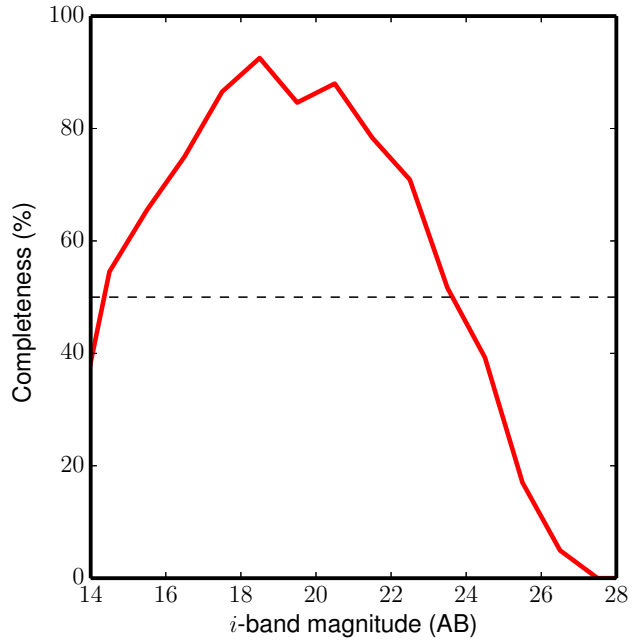}
\caption{Spectroscopic completeness of the \cha \leg survey as a function of $i_{AB}$ (red solid line). 50\% completeness (black dashed line) is also plotted.}\label{fig:spec_complete}
\end{figure}

In Figure \ref{fig:spec_complete} we show the whole \cha \leg survey spectroscopic completeness: the completeness is $\geq$80\% up to a $i$-band magnitude (AB) of 21.5, then there is a linear decline in the completeness value, which is $\simeq$70\% at $i_{AB}$=22.5, $\simeq$50\% at $i_{AB}$=24, finally dropping below 25\% only for sources fainter than $i_{AB}$=25. The relatively low completeness ($\simeq$50\%) at bright magnitudes ($i_{AB}$$<$16) is due to the fact that most of the sources in this magnitude range are stars for which no spectrum was taken. 

\begin{figure}[H]
\centering
\includegraphics[width=0.5\textwidth]{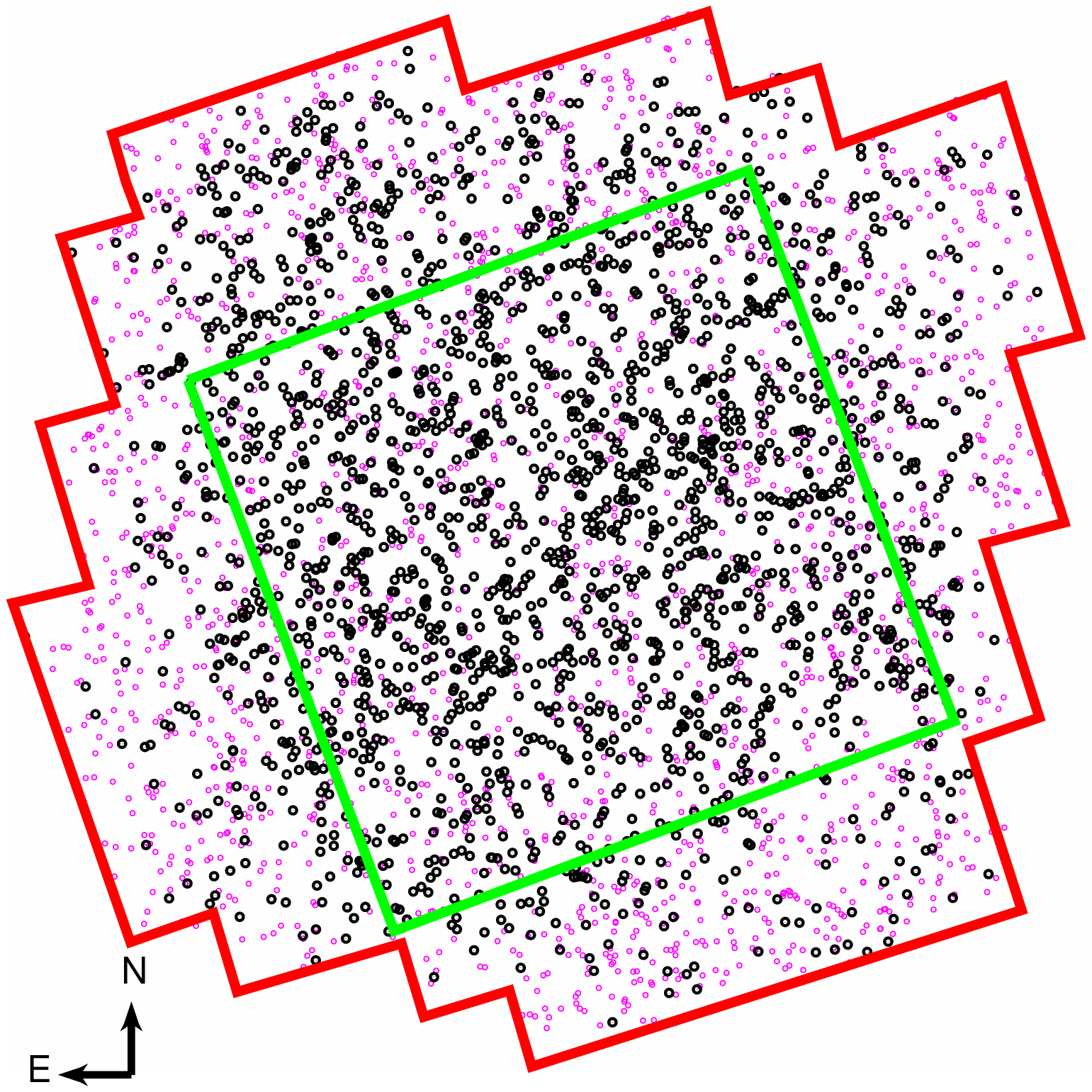}
\caption{Sources with (black circles) and without (magenta circles) spectroscopic redshift in the \cha \leg area (red solid line). The C-COSMOS area is also plotted (green solid line). A significant fraction of sources in the external part of the field has not been spectroscopically followed-up yet.}\label{fig:spec_distrib}
\end{figure}

In Figure \ref{fig:spec_distrib} we show the spatial distribution of the sources with spectroscopic redshift (black circles) on the whole \cha \leg area (red solid polygon): as can be seen, the spectroscopic follow-up of the \cha \leg sources has so far been focused mainly on the central C-COSMOS area (green solid line), while a significant fraction of sources in the external part of the \cha \leg field has not been observed yet. Therefore, the spectroscopic completeness value of the whole survey will easily grow in the coming years, thanks to a dedicated program with \textit{Keck}-DEIMOS (P.I.: G. Hasinger).

\begin{figure*}[!h]
  \begin{minipage}[b]{.5\linewidth}
    \centering
    \includegraphics[width=0.99\textwidth]{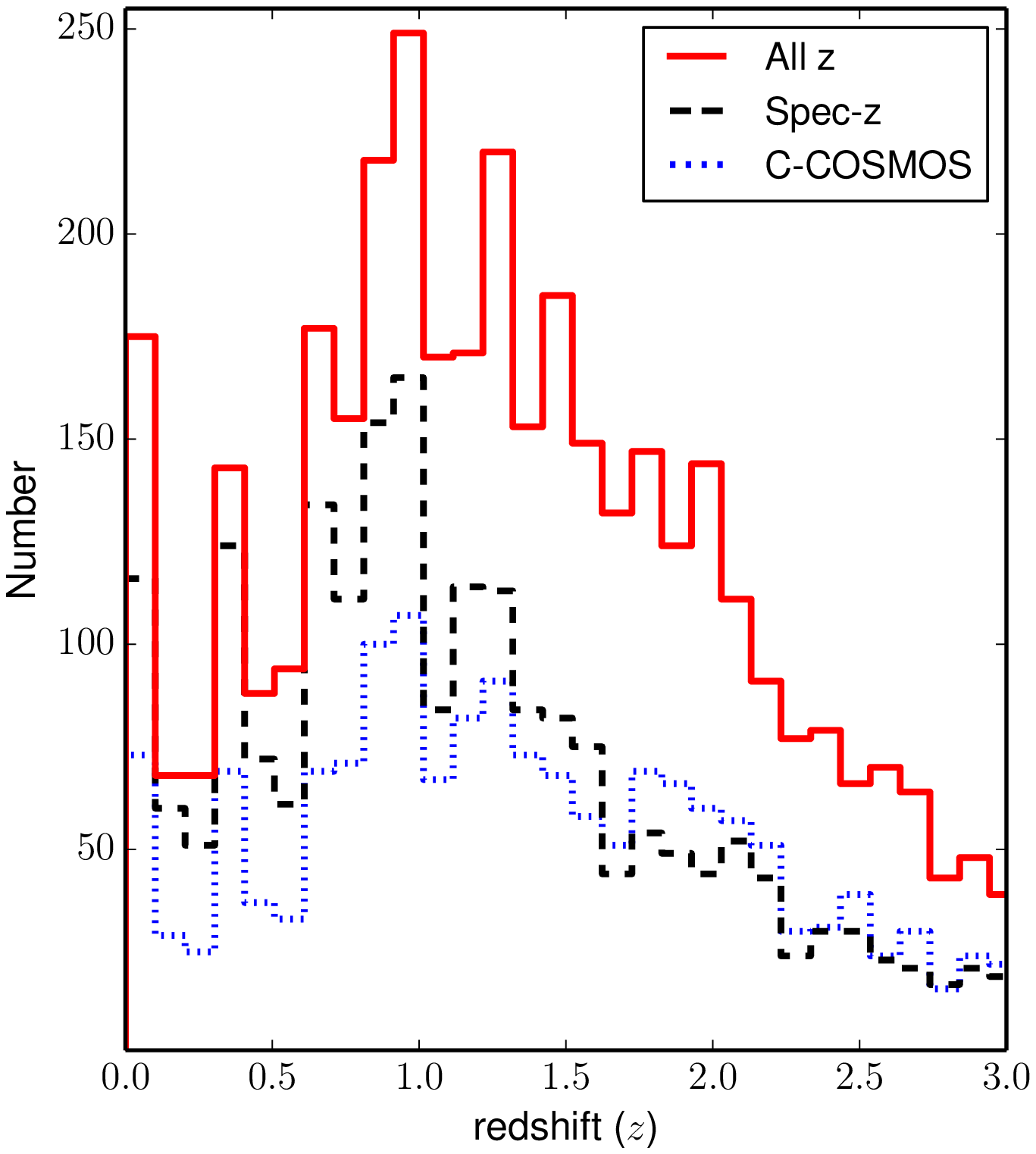}
    \label{subfig-1:i_band}
  \end{minipage}
  \hfill
  \begin{minipage}[b]{.5\linewidth}
    \centering
    \includegraphics[width=1.01\textwidth]{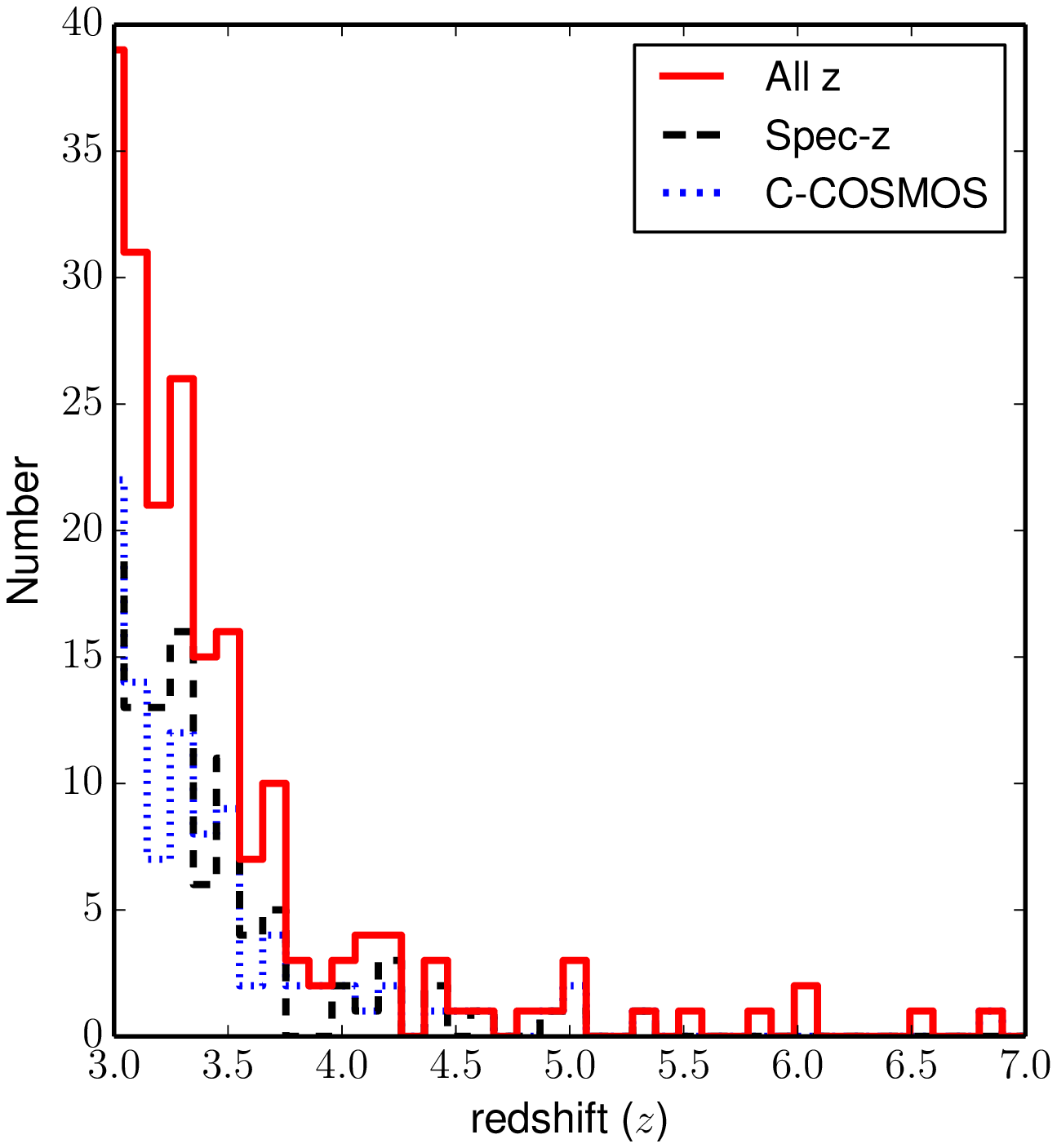}
    \label{subfig-2:k_band}
  \end{minipage}
\caption{Redshift distribution of the whole \cha \leg (red solid line), of the sources with reliable spectroscopic redshifts (black dashed line), and of C-COSMOS spec+photo-z (blue dotted line) for the redshift range $z$=0--3 (left) and $z$=3--7 (right).}
\label{fig:z_histo}
\end{figure*}

The redshift distribution of all \cha \leg sources with a redshift is plotted in Figure \ref{fig:z_histo} (red solid line). The shape of the distribution is consistent with that of C-COSMOS (blue dotted line) and peaks at $z$=1-2. Many spikes are visible in the distribution (see, e.g., $z$$\simeq$1, $z$$\simeq$1.3), and these features are linked to large-scale structures in the COSMOS field (Gilli et al. 2009). 
The evidence of the most prominent spikes linked to the large-scale structures remains also when using only reliable spectroscopic redshifts (black dashed line).

\subsubsection{Sources without optical identification}
80 sources in the whole survey have no optical counterpart and lie inside the optical/IR field of view. We further analyzed these sources, because some of them could be obscured and/or high-redshift AGN (Koekemoer et al. 2004). We visually inspected all these objects, using both X-ray and optical/IR images, and we found that about 50\% of the sources have no optical counterpart because of bad optical imaging, or because the possible counterpart is close to a very bright object (star or extended galaxy) and it is therefore undetected. 

After this visual check, there are still 43 sources without an optical counterpart, but with a $K$-band or 3.6 $\mu$m IRAC counterpart, or with no counterpart at all. 19 of these sources have both a $K$-band and a 3.6 $\mu$m IRAC counterpart, 7 have only a $K$-band counterpart, 7 have only a  3.6 $\mu$m IRAC counterpart and 10 have no counterpart at all. Nine of these sources have no soft--band detection, thus suggesting high obscuration rather than high redshift: seven of these sources have DET\_ML$>$16 either in the full or the hard band, i.e., they are significant at $\geq$5.5$\sigma$; the remaining two sources have DET\_ML$\sim$12 in either the full or the hard band, closer to the survey limit DET\_ML=10.8, and therefore may be spurious X-ray detections. 

\subsection{High redshift sample}
\cha \leg is also the X-ray survey on a single contiguous field with the highest number of high redshift sources: in the whole field there are 174 sources with $z\geq$3 (85 of which have reliable spec-z), 27 sources with $z>$4 (11 with reliable spec-z), 9 sources at $z$$>$5 (2 with reliable spec-z) and 4 sources (3 of which are new, all 4 are photo-z) at  $z>$6. The source with the highest spectroscopic redshift, $z$=5.3, lies in a proto-cluster, where it is also the only X-ray source detected (Capak et al. 2011; Kalfountzou et al. in prep.). A detailed discussion of the sources at $z\geq$3, together with an extended analysis of the space density of the X-ray sources in this redshift range, will be presented in Marchesi et al. (to be submitted). 

\subsection{Spectroscopic and photometric types}\label{sec:spec_phot_type}
We report in Table \ref{tab:spec_type} the characterization of the sources by spectroscopic type (when available) for the new \cha \leg sources and for those in C-COSMOS. In the same table we also show how sources have been divided on the basis of the template which best fits the SED of the sources.

In the whole survey, there are 1770 sources with a reliable spectroscopic redshift and a spectral type information; 722 of these are new sources. Of these 1770  sources, 632 (36\% of the spectroscopic sample with spectral type information) show evidence of at least one broad  (i.e. with FWHM$>$2000 km s$^{-1}$) line in their spectra (BLAGN). There are 1049 sources (59\% of the spectroscopic sample with spectral type information) with only narrow emission lines or absorption lines. These objects are defined as ``non broad-line AGN'' (non-BLAGN). We do not make a further separation between star-forming galaxies and Type 2 AGN on the basis of the source spectra, because the large majority of these sources have low SNR spectra (mainly obtained just to determine the redshift) or are in an observed wavelength range which does not allow to use optical emission line diagnostic diagrams to disentangle in Type 2 AGN and star-forming galaxies.

Finally, the sample contains 89 spectroscopically identified stars (5\% of the spectroscopic sample; see Wright et al. (2010) for a detailed analysis of the stars detected in C-COSMOS). 

It is worth noticing that  $\simeq$58\% of sources in the whole sample are still without spectroscopic type, thus the fractions of different spectral types may be not representative of the complete sample. 

3855 sources (96.0\% of the whole sample) have a photometric SED template information. The largest part (64\%) of these sources are fitted with a non-active galaxy, 9\% are fitted with an obscured AGN template and 23\% by a template with contribution by unobscured AGN. Finally, 121 sources, 3\% of the whole sample, have been identified as stars on the basis of the photometric template. 

We compared the spectroscopic and photometric classifications and we found that 82\% of the sources with BLAGN spectral type have been fitted with an unobscured AGN template, while 97\% of the non-BLAGN are fitted with either a galaxy template (74\%) or with an obscured AGN template (23\%). The lower agreement for BLAGN is not surprising, given that BLAGN SEDs can be contaminated by stellar light; this is particularly true for low-luminosity AGN (Luo et al. 2010; Elvis et al. 2012; Hao et al. 2014). Finally, 84 of the 89 spectroscopically identified stars (94\%) are also photometric stars. As a general assumption, we use the spectroscopic type when available and if not the photometric one. In the following part of the Paper, we refer to BLAGN or unobscured sources as ``Type 1'', and to non-BLAGN or obscured sources as ``Type 2''. 

It is worth noticing that in XMM-COSMOS (B10) there were $\simeq$50\% Type 1 sources and $\simeq$50\% Type 2 sources: \cha \leg reaches a flux limit three times deeper than XMM-COSMOS and therefore samples a larger fraction of obscured objects.

\begin{table*}[!h]
\centering
\scalebox{1.}{
\begin{tabular}{ccccccc}
\hline
\hline
& N$_{new}$ & \%$_{new}$ & N$_{CCosm}$ & \%$_{CCosm}$ & N$_{all}$ & \%$_{all}$\\
\hline
\textbf{Spectroscopic redshifts}&\\ 
\hline
Broad line &       257 & 36 & 375 & 36 & 632 & 36\\
Not broad line & 434 & 60 & 615 & 59 & 1049 & 59\\
Star &                  31 & 4 & 58 & 6 & 89 & 5\\ 
\hline
\textbf{Photometric redshifts}&\\ 
\hline 
Unobscured AGN template & 445 & 21  & 449 & 27 & 894 & 23\\
Obscured AGN template &  261 & 12 & 104 & 6 & 365 & 9\\
Galaxy template & 1398 & 65  & 1077 & 64 & 2475 & 64\\
Star template & 61 & 3 & 60 & 4 & 121 & 3\\ 
\hline
Visually selected star & 8 & & 0 & & 8\\ 
\hline
\hline
\end{tabular}}\caption{Number of X-ray sources divided by spectral or photometric type. N$_{new}$ is the number of sources from the new survey, N$_{CCosm}$ is the number of sources from C-COSMOS and N$_{all}$ is the sum of the previous two values. The fraction is measured on the total number of sources with spectroscopic or SED template best fitting information.}\label{tab:spec_type}
\end{table*}

\subsection{X-ray luminosity}\label{sec:lum_info} 
\begin{figure*}[!h]
  \begin{minipage}[b]{.5\linewidth}
    \centering
    \includegraphics[width=.98\linewidth]{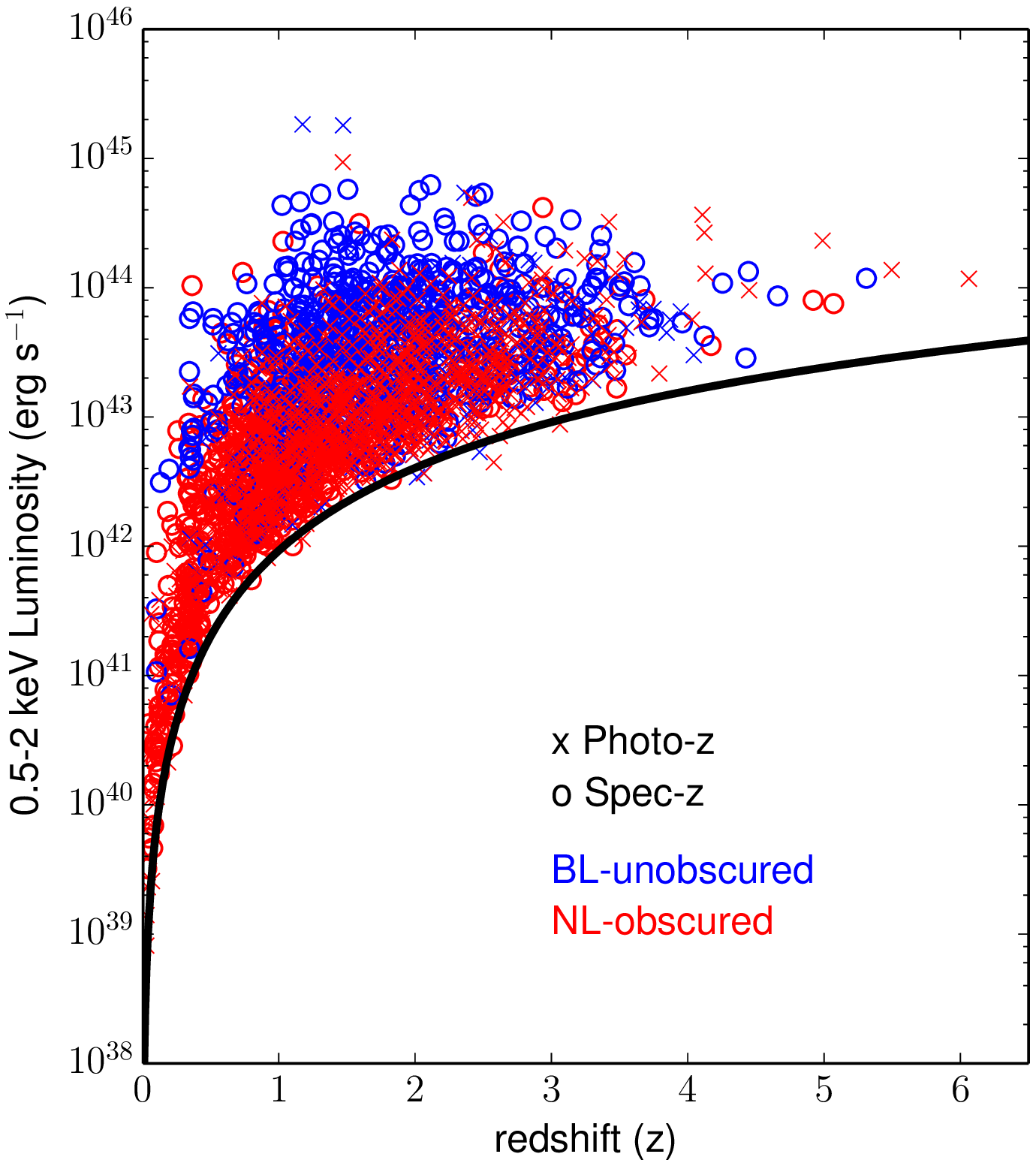}
    \label{subfig-1:i_band}
  \end{minipage}
  \hfill
  \begin{minipage}[b]{.5\linewidth}
    \centering
    \includegraphics[width=.99\linewidth]{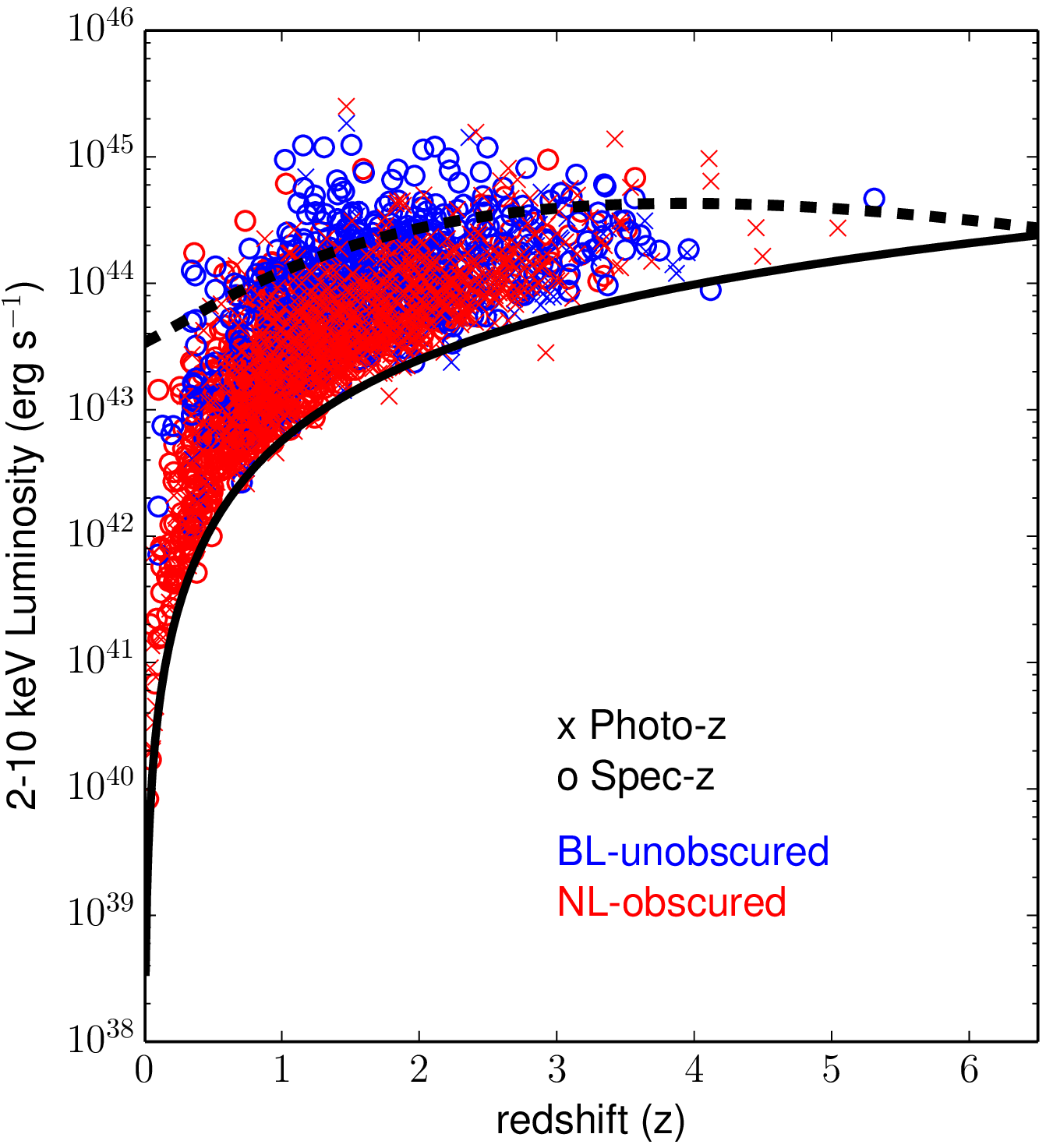}
    \label{subfig-2:k_band}
  \end{minipage}
\caption{Rest-frame luminosity versus redshift in soft (0.5-2 keV, left) and hard (2-10 keV, right). Spectroscopic type (open circles) is plotted when available, otherwise photometric information (cross) is shown. Blue sources are Type 1 AGN; red are Type 2 AGN. We also plotted the survey flux limit (black solid line) and the L$^*$ curve as function of redshift from Aird et al. (2015, black dashed line).}
\label{fig:z_vs_lx}
\end{figure*}

In Figure \ref{fig:z_vs_lx} we show the X-ray luminosity versus redshift, in both soft (left, 2698 out of 4016 sources) and hard (right, 2354 sources) bands, for sources with $z>$0 and DET\_ML$>$10.8. We converted fluxes into luminosities using the best redshift available, i.e. the spectroscopic one when available and the photometric redshift for the remaining sources; we used an X-ray spectral index of $\Gamma$=1.4, to compute $K$-corrected luminosities. We did not apply any obscuration correction. In Figure \ref{fig:z_vs_lx}, right, we also plot the $z$-$L_{2-10 keV}$ curve of the knee of the AGN luminosity function (black dashed line), computed following the Flexible Double Power-Law (FDPL) model from Aird et al. (2015):

\begin{equation}
log L^*(z) = 43.53 + 1.23\times x + 3.35\times x^2 - 4.08\times x^3,
\end{equation} 

where x=log(1+z). As can be seen, we are able to sample with excellent statistics the luminosity range below the knee of the luminosity function, up to redshift $z\simeq$4.   

%
10\% and 3\% of the sources in the soft and hard band, respectively, have luminosities $L_X$$<$10$^{42}$ erg s$^{-1}$, i.e. lower than the threshold which is conventionally used to separate clear AGN from galaxies with no or low nuclear emission, low-luminosity AGN or very obscured AGN (see, e.g., Basu-Zych et al. 2013; Kim \& Fabbiano 2014; Civano et al. 2014; Paggi et al. submitted). This fraction is significantly lower than the fraction of sources that have been fitted with a galaxy SED template (66\% of all the sources). Therefore, the majority of sources fitted with a galaxy template are actually more likely to be obscured AGN rather than normal and starburst galaxies. 

\begin{figure*}[!h]
  \begin{minipage}[b]{.5\linewidth}
    \centering
   \includegraphics[width=1.00\linewidth]{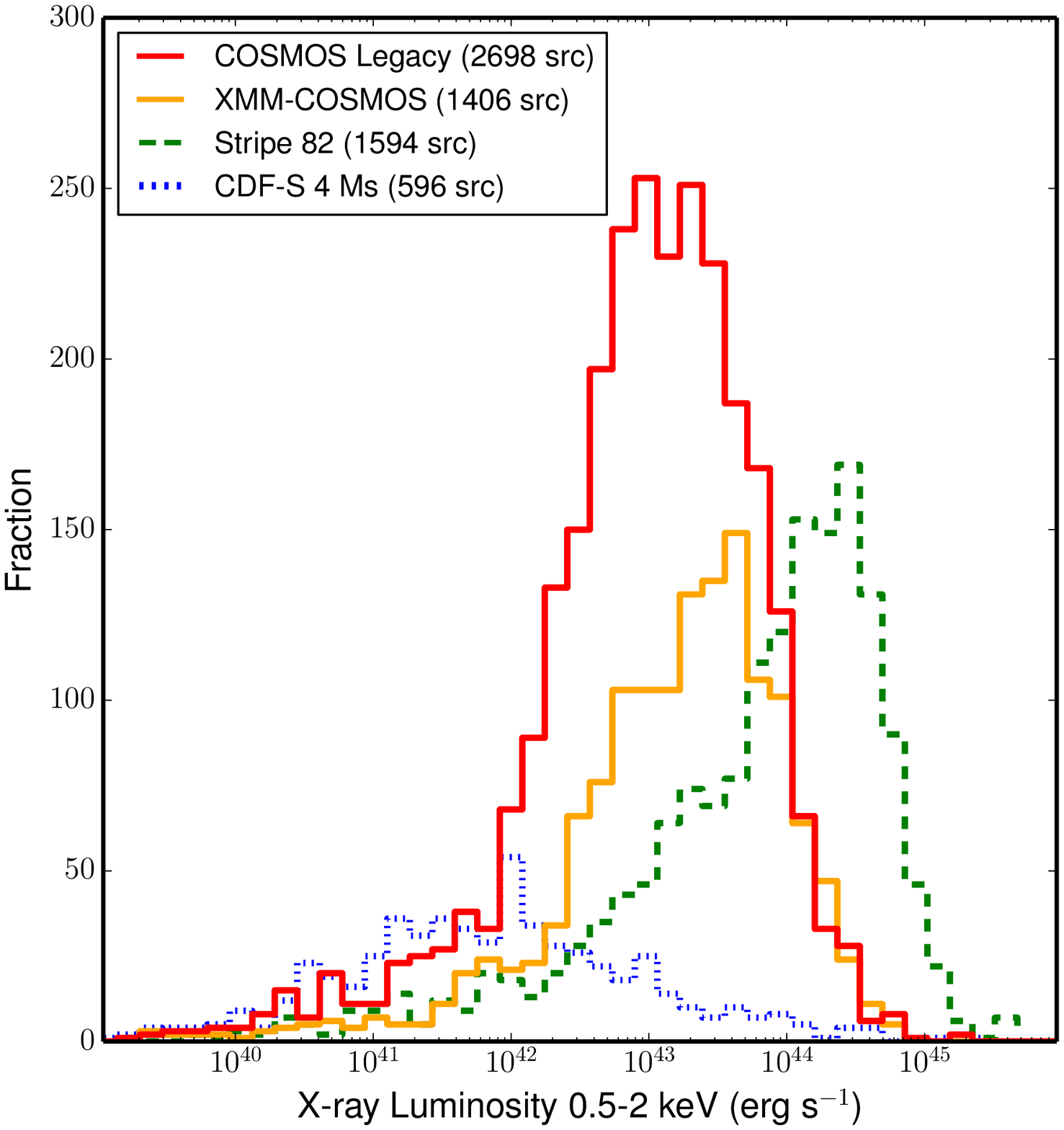}
    \label{subfig-1:i_band}
  \end{minipage}
  \hfill
  \begin{minipage}[b]{.5\linewidth}
    \centering
    \includegraphics[width=1.0\linewidth]{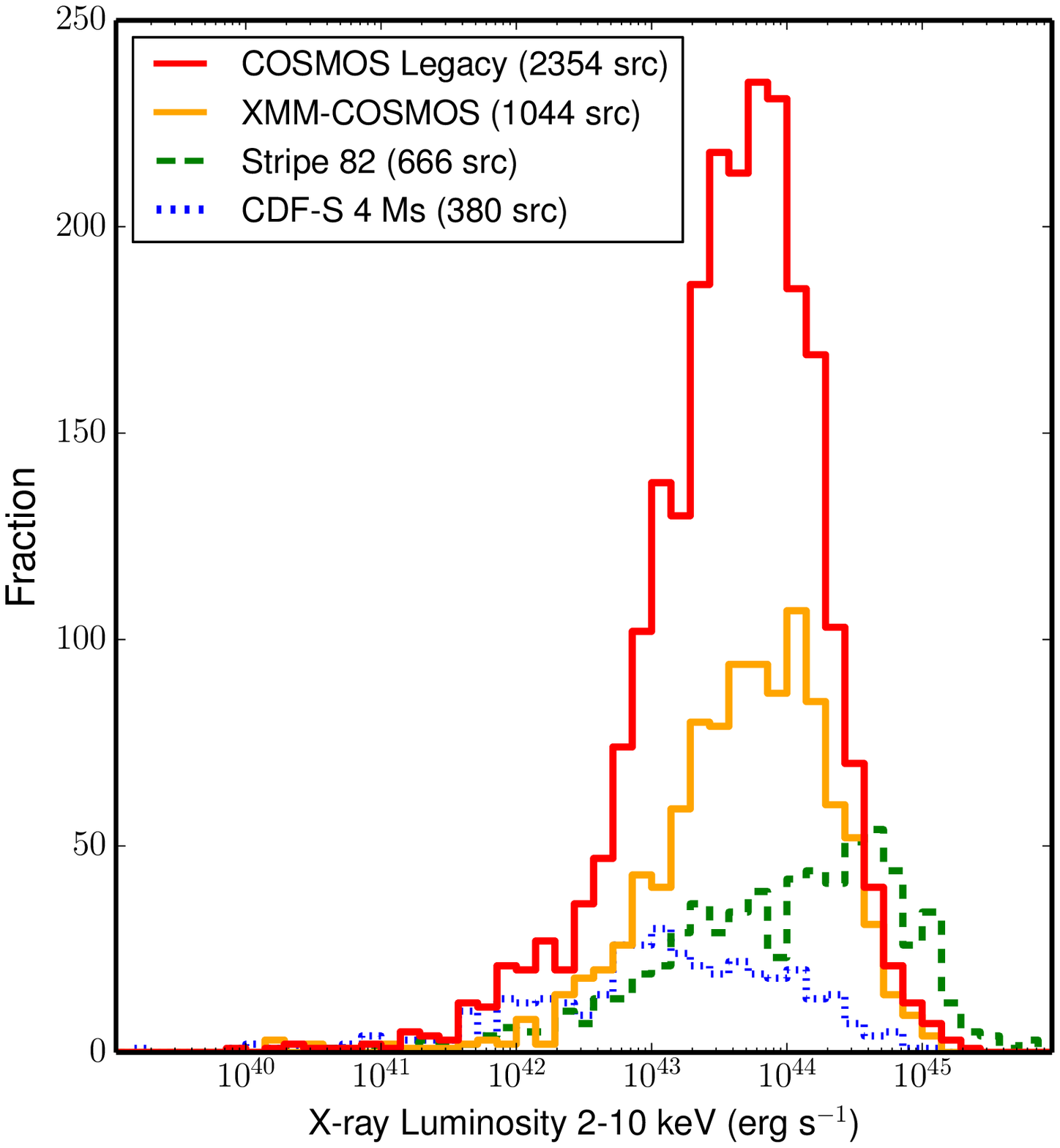}
    \label{subfig-2:k_band}
  \end{minipage}
\caption{Rest-frame luminosity distribution in soft (0.5-2 keV, left) and hard (2-10 keV, right) bands, for all sources in \cha \leg with z$>$0 (spectroscopic or photometric) and DET\_ML$>$10.8 in the given band (red solid line), XMM-COSMOS (orange solid line), CDF-S (blue dotted line) and Stripe 82 (green dashed line).}
\label{fig:histo_lx_w_surveys}
\end{figure*}

\cha \leg luminosity distribution in the soft band peaks at $L_X$$\simeq$3$\times$10$^{43}$ erg s$^{-1}$, and is an excellent bridge between deep pencil beam surveys like CDF-S (Xue et al. 2011), where more than 50\% of the sources have $L_X$$<$10$^{42}$ erg s$^{-1}$ and therefore are more likely to be star-forming galaxies or very obscured AGN, and large area surveys like Stripe 82 (LaMassa et al. 2013), with a luminosity distribution that peaks at $L_X\simeq$ 2 $\times$ 10$^{44}$ erg s$^{-1}$ and whose main goal is to find very bright AGN. This complementarity between different surveys is shown in Figure \ref{fig:histo_lx_w_surveys}, in the left panel. In the same Figure, in the right panel, we show the hard band luminosity distribution, where the peak is at $L_X$$\simeq$ 9 $\times$10$^{43}$ erg s$^{-1}$. 

We also plot in Figure \ref{fig:histo_lx_w_surveys} the luminosity distribution of XMM-COSMOS (B10, orange solid line): as can be seen, XMM-COSMOS already sampled the high luminosity distribution in the COSMOS field, while \cha \leg statistics is significantly better moving towards lower luminosities (i.e.,  $L_X$$\leq$ 5 $\times$ 10$^{43}$ erg s$^{-1}$ in soft and $L_X$$\leq$ 10$^{44}$ erg s$^{-1}$ in hard band, respectively).

\cha \leg covers with an excellent statistics the range of redshift 1$\leq$$z$$\leq$3, i.e. at the peak of the AGN activity and the following period, where the sources span about two orders of magnitude in luminosity (10$^{42.5}$-10$^{44.5}$ erg s$^{-1}$): in redshift range $z$=[1-2] there are 1582 sources, while in the range $z$=[2-3] there are 717 sources. 

\section{The multiwavelength catalog of \textit{Chandra} COSMOS Legacy sources}\label{sec:catalog}
The multiwavelength  catalog of the whole \cha \leg Survey source identifications (i.e., for both the new \cha \leg sources and for the C-COSMOS ones) is available with this Paper, in the COSMOS repository\footnote{http://irsa.ipac.caltech.edu/data/COSMOS/tables/chandra/} and online (in FITS format). The multiwavelength properties reported in the catalog are listed below.
\begin{enumerate}
\item \textit{Column 1}. Source ID. Sources are listed in the same order used in Paper I: first all sources detected in full band, then those detected in soft band only, then those detected in hard band only.
\item \textit{Columns 2-3}. X-ray coordinates of the source, from Paper I catalog.
\item \textit{Columns 4-6}. Maximum likelihood detection (DET\_ML) value in 0.5-7 keV, 0.5-2 and 2-7 keV band, from Paper I catalog.
\item \textit{Columns 7-9}. X-ray fluxes in full, soft and hard bands, from Paper I catalog. Negative fluxes represent upper limits.
\item \textit{Columns 10-12} Hardness ratio and hardness ratio 90\% lower and upper limit, from Paper I catalog.
\item \textit{Column 13}. Identifier number of the optical counterpart from the Ilbert et al. (2009) catalog.
\item \textit{Columns 14-15}. Optical coordinates of the source, from the Ilbert et al. (2009) catalog.
\item \textit{Columns 16-17}. $i$-band magnitude and magnitude error in 3$^{\prime\prime}$ aperture, from the Ilbert et al. (2009) catalog.
\item  \textit{Column 18}. $i$-band magnitude origin: 1 Subaru, 2 CFHT, 3 SDSS, 5 manual photometry
\item  \textit{Column 19}. Identifier number of the $K$-band counterpart from the UltraVISTA catalog from Laigle et al. (submitted).
\item \textit{Columns 20-21}. UltraVISTA $K$-band counterpart coordinates, from the Laigle et al. (submitted) catalog.
\item \textit{Columns 22-23}. UltraVISTA $K$-band magnitude and magnitude error in 3$^{\prime\prime}$ aperture, from the Laigle et al. (submitted) catalog.
\item  \textit{Column 24}. Identifier number of the $K$-band counterpart from the CFHT catalog from Ilbert et al. (2009) catalog.
\item \textit{Columns 25-26}. CFHT $K$-band counterpart coordinates, from the Ilbert et al. (2009) catalog.
\item \textit{Columns 27-28}. CFHT $K$-band magnitude and magnitude error in 3$^{\prime\prime}$ aperture, from the Ilbert et al. (2009) catalog.
\item \textit{Column 29-30}. Coordinates of the 3.6 $\mu$m counterpart from the Sanders catalog.
\item \textit{Column 31-32}. 3.6 $\mu$m flux ($\mu$Jy) and flux error in 1.9$^{\prime\prime}$ aperture, from the Sanders catalog. To convert to total flux, the standard factor suggested in the IRAC user guide has to be applied (dividing by 0.765).
\item \textit{Column 33-34}. Coordinates of the 3.6 $\mu$m counterpart from the SPLASH catalog.
\item \textit{Column 35-36}. 3.6 $\mu$m flux ($\mu$Jy) and flux error in 1.9$^{\prime\prime}$ aperture, from the SPLASH catalog. To convert to total flux, the standard factor suggested in the IRAC user guide has to be applied (dividing by 0.765).
\item \textit{Column 37}. Final identification flag:  1= secure, 10= ambiguous, 100= subthreshold, -99= unidentified
\item \textit{Column 38}. Star flag: 1= spectroscopically confirmed star, 10= photometric star, 100= visually identified star.
\item \textit{Column 39}. Best redshift available. This is the spectroscopic redshift if the spectroscopic redshift quality flag is Qg$\geq$1.5 (see below) and the photometric redshift otherwise.
\item \textit{Column 40}. Spectroscopic redshift.
\item \textit{Column 41}. Spectroscopic redshift origin.
\item \textit{Column 42}. Spectroscopic redshift quality. 2= ``secure'' redshift, spectroscopic reliability $>$99.5\%, 1.5= ``reliable'' redshift, spectroscopic reliability $<$99.5\% but there is a photometric redshift such that $\frac{\Delta z}{1+z_{spec}}$$<$0.1, 1= ``not reliable'' redshift, spectroscopic reliability $<$99.5\% and there is a photo-z such that $\frac{\Delta z}{1+z_{spec}}$$>$0.1.
\item \textit{Column 43}. Spectroscopic identification. 1=BLAGN, 2=non-BLAGN, 0=star.
\item \textit{Column 44}. Photometric redshift from Salvato et al. (in preparation).
\item \textit{Column 45}. Photometric identification from SED fitting (1=unobscured, 2=obscured, 3=galaxy, 5=star).
\item \textit{Column 46}. Identifier number of the \textit{XMM-COSMOS} counterpart, from the Cappelluti et al. (2009) catalog.
\item \textit{Column 47}. Luminosity distance (in Mpc).
\item \textit{Columns 48-50}. Rest-frame luminosity, in 0.5-10 keV, 0.5-2 keV and 2-10 keV bands, obtained assuming an X-ray spectral index $\Gamma$=1.4.
\item \textit{Column 51}. Intrinsic neutral hydrogen ($N_H$) column density, estimated using the best redshift available and the hardness ratio from Paper I catalog, assuming an X-ray spectral index $\Gamma$=1.8.
\item \textit{Columns 52-54}. Luminosity absorption correction, in 0.5-10 keV, 0.5-2 keV and 2-10 keV bands, obtained assuming the intrinsic $N_H$ reported in \textit{Column 47} and a power-law with spectral index $\Gamma$=1.8.
\item \textit{Column 55}. Lower limit on intrinsic $N_H$ column density, estimated using the best redshift available and the hardness ratio lower limit from Paper I catalog, assuming an X-ray spectral index $\Gamma$=1.8.
\item \textit{Columns 56-58}. Luminosity absorption correction, in 0.5-10 keV, 0.5-2 keV and 2-10 keV bands, obtained assuming the intrinsic $N_H$ reported in \textit{Column 51} and a power-law with spectral index $\Gamma$=1.8.
\item \textit{Column 59}. Upper limit on intrinsic $N_H$ column density, estimated using the best redshift available and the hardness ratio upper limit from Paper I catalog, assuming an X-ray spectral index $\Gamma$=1.8.
\item \textit{Columns 60-62}. Luminosity absorption correction, in 0.5-10 keV, 0.5-2 keV and 2-10 keV bands, obtained assuming the intrinsic $N_H$ reported in \textit{Column 55} and a power-law with spectral index $\Gamma$=1.8.
\end{enumerate}

\section{X-ray, optical and infrared properties}\label{sec:opt_ir}

\subsection{Redshift Evolution of Hardness Ratio}\label{sec:hr}
Through unbinned statistics and careful background modelization, the minimum number of counts required for the X-ray spectral analysis is set only by the maximum relative error that one wants to allow. However, assuming a threshold of 70 net counts (Lanzuisi et al. 2013), there are only $\simeq$950 of the 4016 sources in our survey (i.e. $\simeq$24\%) that fulfill this requirement. Nevertheless, it is possible to use the Bayesian Estimation of Hardness Ratios (BEHR) method (Park et al. 2006) to derive a rough estimation of the X-ray spectral shape and therefore of the source nuclear obscuration. The hardness ratio (HR) of the source is defined as the ratio $\frac{H-S}{H+S}$, where H and S are the net counts of the source in the hard (2-7 keV) and in the soft (0.5-2 keV) band, respectively: an extended description of the procedure adopted to compute HR is reported in Paper I. BEHR is particularly effective in the low count regime, because it does not need a detection in both bands to work and it runs Markov chain Monte Carlo calculation to compute errors.

To separate unobscured and obscured sources, we adopted a redshift dependent HR threshold (HR$_{th}$), computed assuming a typical obscured AGN spectrum, with a power-law with $\Gamma$=1.4: consequently, we consider sources with HR$>$HR$_{th}$ as obscured. For sources with no redshift information, we used HR$_{th}$=-0.2, i.e. the mean HR value of our redshift-dependent curve. 1993 sources in \cha \leg ($\simeq$49.6\% of the whole sample) have HR$>$HR$_{th}$, including both nominal values and 90\% significance lower limits. We point out that such a value should be treated as a lower limit on the obscuration of the AGN population in COSMOS, particularly for those sources at high redshift and low-luminosity. There are in fact two main caveats involved in the use of the HR threshold: ($i$) the soft appearance of a fraction of Compton Thick sources at high redshift (Brightman et al. 2014), where we observe the intrinsic hard band emission in the soft band; ($ii$) a fraction of more obscured sources (at a given intrinsic flux) have flux below the flux limit of the survey and is therefore missed (Wilkes et al. 2013).

\begin{figure}[H]
\centering
\includegraphics[width=0.5\textwidth]{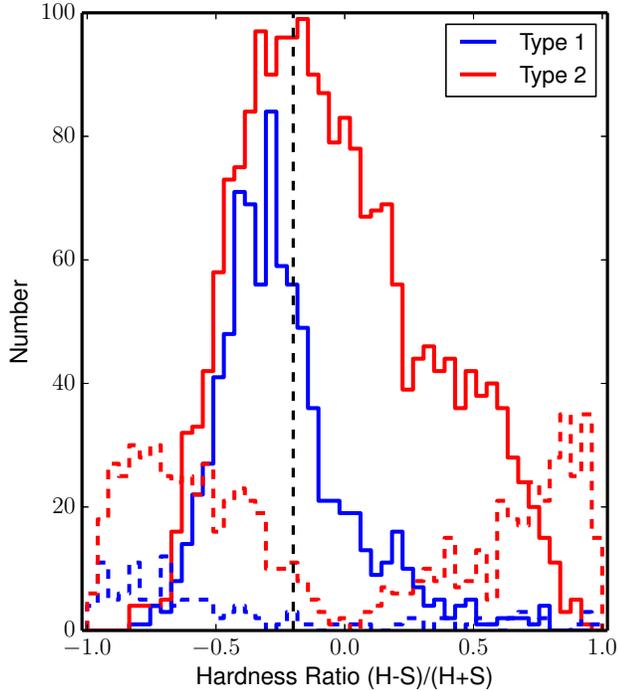}
\caption{Average HR distribution for optically classified Type 1 (blue) and Type 2 (red) sources. Upper (towards the left) and lower (towards the right) limits are plotted as dashed lines. The black dashed line at HR=-0.2 marks the average HR$_{th}$, computed assuming a typical obscured AGN spectrum, with a power-law with $\Gamma$=1.4.}\label{fig:histo_hr_w_type}
\end{figure}

In Figure \ref{fig:histo_hr_w_type}, we show the HR distribution for optically classified Type 1 (blue) and Type 2 (red) sources: spectral types are used when available, and the best-fit SED template model for the remaining sources. The mean (median) HR  is HR=-0.26$\pm$0.32 (-0.3) for Type 1 sources and HR=-0.03$\pm$0.46 (-0.10) for Type 2 sources, taking in account in the computation also the 371
lower limits and the 616 upper limits (shown in Figure \ref{fig:histo_hr_w_type} as dashed lines). The hypothesis that the two distributions are actually the same is rejected on the basis of a Kolmogorov-Smirnov (KS) test, with a probability $>$99.998\%. A similar result was already shown in B10 in XMM-COSMOS: we found that the values do not change significantly if we use only a subsample with flux f$_{0.5-10}<$ 5$\times$10$^{-15}$ erg s$^{-1}$ cm$^{-2}$, i.e. in the range where \cha \leg statistics is significantly larger than the XMM-COSMOS one, and are therefore not dominated by the brightest sources.

\begin{figure}[H]
\centering
\includegraphics[width=0.5\textwidth]{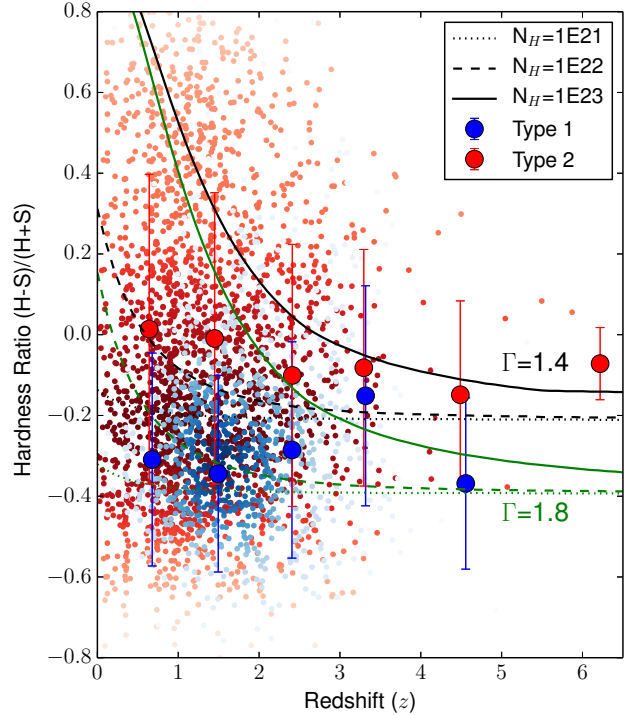}
\caption{HR evolution with redshift for optically classified Type 1 (blue) and Type 2 (red) sources. The error bars represent the 68\% dispersion. Three curves of different $N_H$ (10$^{21}$ cm$^{-2}$, dotted line, 10$^{22}$ cm$^{-2}$, dashed line, 10$^{23}$ cm$^{-2}$, solid line) are also plotted for comparison, obtained assuming a power-law spectrum with $\Gamma$=1.4 (black) and $\Gamma$=1.8 (green). Single values for each source with significant HR are plotted in the background (darker scale color indicates higher source density).}\label{fig:histo_hr_vs_z_w_type}
\end{figure}

Finally, we studied the behavior with redshift of the HR: we show the result in Figure \ref{fig:histo_hr_vs_z_w_type}, where once again we divide our sample in Type 1 (blue) and Type 2 (red) sources, on the basis of the optical classification. We also show three curves of different column density ($N_H$=10$^{21}$,10$^{22}$ and 10$^{23}$ cm$^{-2}$, dotted, dashed and solid line, respectively), obtained assuming a power-law spectrum with $\Gamma$=1.4 (black) and $\Gamma$=1.8 (green). As can be seen, the average HR of Type 2 lies above the $N_H$=10$^{22}$ cm$^{-2}$ curve at all redshifts, regardless of the assumed $\Gamma$, while the average HR of Type 1 sources is generally below the $N_H$=10$^{21}$ cm$^{-2}$ curve computed assuming $\Gamma$=1.4. However, the large dispersion in the HR distribution, at any redshift, does not allow to claim that the optically classified Type 1 and Type 2 sources lie in two different regions of the HR versus redshift diagram. Such a dispersion (in Figure \ref{fig:histo_hr_vs_z_w_type} we show the 68\% dispersion) is particularly large for Type 2 sources ($\sigma>$0.3 at $z<$3), where it is at least partially due to the fact that a significant fraction of sources with a galaxy best-fit SED template are actually objects where the galaxy optical contribution is dominant, and it is therefore not possible to correctly classify the AGN; in the X-ray, instead, the AGN contribution is almost unbiased even at the \cha \leg flux limit. We discuss further the different information obtained using the HR as an obscuration indicator, instead of the optical classification, in Section \ref{sec:lx_vs_obscured}.

\subsubsection{Intrinsic $N_H$ and de-absorbed luminosity estimation}\label{sec:nh_estimate}
To estimate the intrinsic $N_H$ of the sources in our sample we used the best available redshift and the HR of each source, using a sample of redshift vs HR curves like those shown in Figure \ref{fig:histo_hr_vs_z_w_type}. These curves have been obtained assuming an X-ray spectral power-law with slope $\Gamma$=1.8. We did not estimate a $N_H$ value for sources without a reliable redshift. After estimating $N_H$, we compute the intrinsic absorption correction $k_{abs}$=$f_{abs}$/$f_{int}$, where $f_{int}$ and $f_{abs}$  are the intrinsic and absorbed fluxes in a given band, respectively.
Finally, we repeated the whole procedure using the HR lower and upper limits, therefore estimating upper and lower limits on the $N_H$.

We compared our $N_H$ estimation with those from Lanzuisi et al. (2013) on the subsample of 388 sources with more than 70 net counts in C-COSMOS, and we found a general good agreement. The sample can be divided as follows:
\begin{enumerate}
\item About 56\% of the sources have only an upper limit on $N_H$ in both our sample and the Lanzuisi et al. (2013) one, and for $\simeq$95\% of these sources this upper limit is $<$10$^{22}$ cm$^{-2}$ in both samples. 
\item $\simeq$18\% of the sources have a significant $N_H$ value in both samples: for these sources the agreement between $N_H$ estimations is generally good, with a mean (median) ratio $r$=0.95 (0.88) between the Lanzuisi et al. (2013) $N_H$ estimation and ours. We did not find a significantly change in the ratio distribution at different fluxes. 
\item  $\simeq$26\% of the sources have a significant detection in one sample and only an upper limit in the other, and more than 90\% of the sources in this last subsample have actually a significant detection in Lanzuisi et al. (2013) and only an upper limit in our sample. This discrepancy can be explained with the better accuracy that the spectral analysis provides with respect to the HR-based estimation: it is also worth noticing that the majority of our upper limits are located within the 1$\sigma$ uncertainty provided by Lanzuisi et al. (2013). Finally, the mean (median) redshift of this sample, z=1.60$\pm$0.73 (z=1.52), is slightly higher, although consistent within the errors, than the mean (median) redshift for the sources with an upper limit in both samples or a significant detection in both samples, z=1.33$\pm$0.72 (z=1.21).
\end{enumerate}

A spectral analysis of the $\simeq$950 sources with more than 70 net counts in the whole \cha \leg survey (included the 388 sources already analyzed) has already been planned (Lanzuisi et al. in prep.); moreover, the excellent \cha \leg statistics will allow to perform stacked spectral analysis of sources with similar properties (e.g. optically classified Type 1 and Type 2 AGN), and therefore compute average $N_H$.

\subsection{X-ray to Optical Flux Ratio}\label{sec:x_vs_opt}
Since the beginning of X-ray surveys, a typical way to characterize different types of X-ray sources has been the X-ray to optical flux ratio (hereafter X/O), which is a simple first estimator of the source classification (Tananbaum et al. 1979; Maccacaro et al. 1988), 

\begin{equation}\label{eq:xo}
X/O = log(f_X/f_{opt})=log(f_X) + C + m_{opt}/2.5,
\end{equation}

where $f_X$ is the X-ray flux in a given band, $m_{opt}$ is the magnitude in the chosen optical band and $C$ is a constant related to the filter used in the optical observations and the band in which the X-ray flux is measured. The magnitude used in this equation is usually the $i$ or $r$-band one (see Brandt \& Hasinger 2005). The relation was first used in the soft X-ray band: in this band, the largest part of bright spectroscopically identified AGN, both BLAGN and non-BLAGN, lie in the region X/O=0$\pm$1 (e.g. Schmidt et al. 1998; Stocke et al. 1991; Lehmann et al. 2001), hereafter defined as the ``soft locus''. \cha and \xmm studies extended this relation to harder bands (Hornschemeier et al. 2001, 2-8 keV; Alexander et al. 2001, 2-8 keV; Giacconi et al. 2002, 0.5-10 keV; Fiore et al. 2003, 2-10 keV; Della Ceca et al. 2004, 4.5-7.5 keV; Cocchia et al. 2007, 2-10 keV). The trend (i.e. the existence of a ``hard locus'', a general correlation between X-ray and optical fluxes) was confirmed at bright fluxes also in these bands, but with a non negligible scatter around the median values, both in soft and hard band, at lower fluxes (Brandt \& Hasinger 2005). 

This scatter is linked to different types of objects: obscured AGN ($N_H$$>$10$^{22}$ cm$^{-2}$) generally lie in the region with X/O$>$1 (Fiore et al. 2003; Perola et al. 2004; Civano et al. 2005; B10); normal, low X-ray flux galaxies have X/O$<$-2 (Xue et al. 2011). Finally, a third class of objects is defined, formed by unobscured X-ray Bright, Optically Normal Galaxies (XBONGs, see Elvis et al. 1981, Comastri et al. 2002; Civano et al. 2007; Trump et al. 2009b). These peculiar sources were named extreme or ``unconventional'' (Comastri et al. 2003; Mignoli et al. 2004) or ``elusive'' (Maiolino et al. 2003), especially when X/O is defined in the hard X-ray band.

\begin{figure*}[!h]
\centering
\includegraphics[width=1.\textwidth]{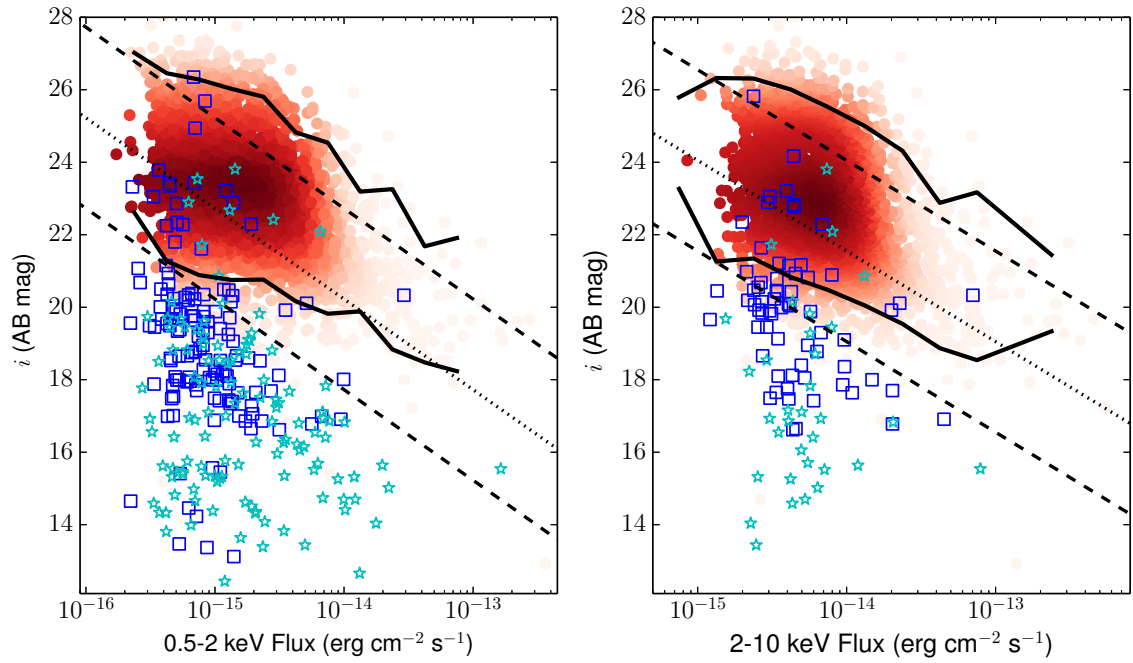}
\caption{X-ray flux (soft on the left, hard on the right) versus $i$-band total (aperture corrected) magnitude, for all X-ray sources with an $i$-band counterpart. The black dashed lines define the so-called ``soft locus'' and ``hard locus'' of AGN along the correlation X/O=0$\pm$1. Red circles are AGN ($L_X$$>$10$^{42}$ erg s$^{-1}$, darker scale color indicates higher source density), blue squares are sources with $L_X$$<$10$^{42}$ erg s$^{-1}$ and cyan stars are stars. Black solid lines represent the region including 90\% of the \cha \leg AGN population.}\label{fig:x_vs_i_w_perc}
\end{figure*}

We studied the X-ray flux versus optical magnitude relation using the whole \cha \leg dataset, in order to put better constraints on it, especially at the X-ray faint end, where our sample is twice as large as the C-COSMOS one. In Figure \ref{fig:x_vs_i_w_perc} we show the relation between the $i$-band magnitude and the X-ray flux in both soft (left) and hard (right) bands for the whole \cha \leg survey: our sample comprises only sources with $i$-band magnitude, DET\_ML$>$10.8 in the given X-ray band and with redshift available, and contains 2777 sources in the soft band and 2353 sources in the hard band. The ``soft locus'' and the ``hard locus'' are also plotted, using a constant $C(i)$=5.91 in the soft band and $C(i)$=5.44 in the hard band. The constant has been computed on the basis of the $i$-band filters width, for all the filters in COSMOS (Subaru, CFHT and SDSS). 

We studied the $i$-band-X-ray flux relation of the whole \cha \leg by dividing our sample in three different subsamples: ($i$) candidate AGN population (red circles), i.e., sources with $L_X$$>$10$^{42}$ erg s$^{-1}$ in full band (or in the soft or hard band if the source was not detected in the full band; 2523 in the soft band and 2253 in the hard band); ($ii$) low-luminosity sources (blue squares, $\simeq$5\% and of $\simeq$3\% the soft and hard samples, respectively: 135 sources in the soft band and 67 in the hard band), i.e., objects with $L_X$$<$10$^{42}$ erg s$^{-1}$; ($iii$) stars (cyan stars, 119 in the soft band and 33 in the hard band). 

A significant fraction of sources with $L_X$$>$10$^{42}$ erg s$^{-1}$  ($\simeq$19\%) lie outside both the soft locus and the hard locus. We then computed the 90\% width of the X/O distribution, i.e., tracing the 5\% lower percentile and the 95\% upper percentile of the $i$-band distribution of the AGN population. To do so, we divided the sources in X-ray flux bins of width 0.25 dex: the results are shown as black solid lines in Figure \ref{fig:x_vs_i_w_perc}. We call this the \cha \leg locus. 

The \cha \leg locus is shifted to fainter optical magnitudes relative to both the soft and hard locus by $\Delta$(X/O)$\simeq$0.3-0.5 in both bands, and  does not change significantly over 1.5 dex in flux. The \cha \leg locus is consistent with that of C12 at any flux and is consistent with the X/O being defined with soft X-ray selected sources, which are usually bright both in the optical band and in the X-rays. 

The majority of stars and candidate low luminosity AGN or non active galaxies (i.e. sources with $L_X$$<$10$^{42}$ erg s$^{-1}$) lie in the region of Figure \ref{fig:x_vs_i_w_perc} at low X-ray fluxes and bright optical magnitudes. However, there is a fraction of sources with low $L_X$ which show X-ray to optical properties consistent with those of sources with $L_X$$>$10$^{42}$ erg s$^{-1}$: 20 of 135 sources with $L_X$$<$10$^{42}$ erg s$^{-1}$  (15\%) lie inside the soft \cha \leg locus, while 19 of 67 (28\%) lie inside the hard \cha \leg locus. The fraction is considerably higher in the hard band, where it is more likely to observe obscured AGN at low-medium redshift. 
The HR distribution of the sources inside and outside the \cha \leg locus is consistent in the soft band, where the sources inside the locus have mean (median) HR=--0.15$\pm$0.26 (HR=--0.14), while the sources outside the locus have mean (median) HR=--0.17$\pm$0.34 (HR=--0.19). In the hard band the sources inside the locus have mean (median) HR=--0.11$\pm$0.33 (HR=--0.14), slightly softer, although consistent within the errors, than those of the sources outside the locus, which have mean (median) HR=0.12$\pm$0.38 (HR=0.08). However, even in this last case the hypothesis that the two distributions are actually the same can not be rejected on the basis of a KS-test (p-value=0.06).
A more accurate analysis of this subsample of candidate obscured AGN is beyond the purpose of this work and requires an extended analysis of parameters like those derived from a morphological analysis of the sources (see Xue et al. 2011; Ranalli et al. 2012). We are also planning a detailed spectral analysis of the low luminosity sources (Lanzuisi et al. in preparation), to determine the average spectral slope $\Gamma$ and the intrinsic absorption $N_H$ of the sources inside and outside the AGN locus.

\begin{figure}[H]
\centering
\includegraphics[width=0.5\textwidth]{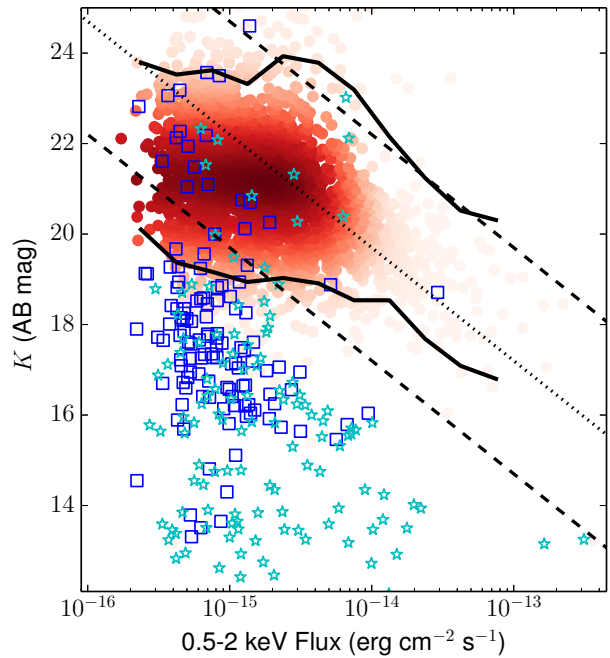}
\caption{Soft X-ray flux versus $K$-band total (aperture corrected) magnitude, for all X-ray sources with a $K$-band counterpart. The black dashed lines define the so-called ``soft locus'' of AGN along the correlation X/O=0$\pm$1. Red circles are AGN ($L_X$$>$10$^{42}$ erg s$^{-1}$, darker scale color indicates higher source density), blue squares are sources with $L_X$$<$10$^{42}$ erg s$^{-1}$ and cyan stars are stars. Black solid lines represent the region including 90\% of the \cha \leg AGN population.}\label{fig:x_vs_k}
\end{figure}

We also studied the trend with X-ray soft flux of the $K$ and 3.6 $\mu$m magnitudes: the two samples contain 2824 and 2868 sources with $L_X$$>$10$^{42}$ erg s$^{-1}$, respectively. Here the soft locus has been computed with Equation \ref{eq:xo}, using constants $C$=6.86 and $C$=7.34 for the $K$ and 3.6 $\mu$m bands, respectively. We computed again also the region which contains 90\% of the AGN population and we found that this region is smaller($\sim$ 1 mag) than in the $i$-band (we show the $K$-band relation in Figure \ref{fig:x_vs_k}). This narrower relation suggests that the relation of $K$ and 3.6 $\mu$m magnitudes with the X-ray flux is stronger than that of the $i$-band one, an evidence which is also reflected in the higher identification rates for $K$ and 3.6 $\mu$m counterparts. Such a result could be mainly linked to the lower contribution of the nuclear extinction at near-infrared wavelengths (Mainieri et al. 2002; Brusa et al. 2005).

\subsection{X/O--hard band luminosity relation with spectroscopic and photometric classification}

\begin{figure}[H]
\centering
\includegraphics[width=0.5\textwidth]{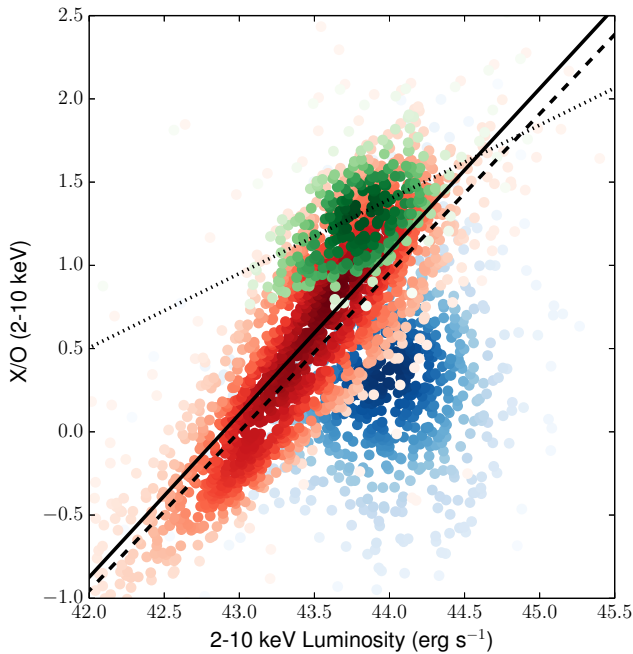}
\caption{X/O versus hard band luminosity, rest frame, for \cha \leg sources with $L_X$$>$10$^{42}$ erg s$^{-1}$ in the 2-10 keV band. Blue sources are Type 1 AGN; red are Type 2 AGN. Sources with $i$$>$25 are plotted in green. Darker scale colours indicate higher source density. The best fit relation for all non BLAGN or obscured AGN and galaxy dominated objects with $L_X$$>$10$^{42}$ erg s$^{-1}$(black solid line), for those with $f_X$(2-10 keV)$>$8$\times$10$^{-15}$ erg s$^{-1}$ cm$^{-2}$ (black dashed line) and for those with $i$$>$25 (black dotted line) are also plotted.}
\label{fig:xo_vs_lx}
\end{figure}

In Figure \ref{fig:xo_vs_lx} we show X/O versus the hard band X-ray luminosity for the 2243 sources with a significant detection in the hard band, with optical counterpart, with spectroscopic or photometric classification and with $L_X$$>$10$^{42}$ erg s$^{-1}$ in the 2-10 keV hard band. Fiore et al. (2003) showed the existence of a linear correlation between X/O and the hard X-ray luminosities for Type 2 AGN. Such a correlation is due to the fact that extinction strongly reduces the nuclear UV/optical emission (where the only remaining contribution is from the host galaxy), but it is instead not heavily attenuated in the 2-10 keV band, at least for sources with $N_H$$<$10$^{24}$ cm$^{-2}$.

In the \cha \leg sample,  Type 2 AGN (red) show a clear linear trend over more than three orders of magnitude, with slope 0.98$\pm$0.02 (black solid line) and correlation coefficient $\rho$=0.81, with p-value=0. This subsample consists of 1551 sources, out of which 646 are spectroscopic Type 2 AGN, 59 are sources with photometric redshifts and SED fitted with an obscured AGN template, and the remaining 846 sources have photo-z and SED best fitted with a galaxy template. On the other hand, unobscured AGN (blue, 692 sources, out of which 518 with spectroscopic information and the remaining 174 with only photometric information) do not show a clear trend between hard X-ray luminosity and X/O: Type 1 AGN are on average 0.5 dex more luminous than non-Type 1 AGN (97\% of the Type 1 sources have $L_X$$>$10$^{43}$ erg s$^{-1}$), but there are many sources with X/O$<$0 even at high X-ray luminosity. This is an expected result, because BLAGN have by definition low obscuration, so the optical flux is higher than in Type 2 AGN, at any X-ray flux. 

We then tested this relation only for the 513 Type 2 sources with $F_X$$>$8$\times$10$^{-15}$ erg s$^{-1}$ cm$^{-2}$ in the hard band, i.e., the flux limit of HELLAS2XMM (Fiore et al. 2003), where the trend between X/O and L$_X$(2-10 keV) was first reported. For this subsample, a linear relation (black dashed line) still exists, with slope 0.95$\pm$0.03 and correlation coefficient $\rho$=0.81, with p-value=0. 

At the faint end of the {Type 2} optical counterparts ($i>$25, 284 sources, green) we instead found a considerably weaker trend (black dotted line), with slope 0.45$\pm$0.04 and correlation coefficient $\rho$=0.54, with p-value=0, confirming that the relation between X/O and $L_X$ becomes flatter, if not totally disappears, moving to faint magnitudes (Barger et al. 2005; Civano et al. 2005). This trend could be partially explained with a selection effect, but when we selected other optical magnitude ranges we found that the relation still exists, even if less steep (for example, for $i$=[21-23] the relation has slope 0.69$\pm$0.04 and $\rho$=0.82, while for $i$=[22-24] the relation has slope 0.65$\pm$0.04 and $\rho$=0.78, with p-value=0 in both cases). However, it is also worth noticing that in the $i>$25 subsample about 90\% of the sources have only a photometric redshift available, and optically faint objects have less reliable photo-z (Salvato et al. 2009; Ilbert et al. 2010). Consequently, the flattening of the slope could be partially caused by an average over-estimation of the faint objects redshifts and consequently luminosities (and star formation rates). On the other hand, the trend does not change significantly (slope 0.48$\pm$0.10 and correlation coefficient $\rho$=0.61, with p-value=5$\times$10$^{-5}$), with respect to the whole Type 2 with $i>$25 subsample,  when we select only the 30 sources with $i>$25 and reliable spec-z.

We tested several parameters to determine potential physical causes of the less significant correlation in the optically faint subsample. 
\begin{enumerate}
\item There is no difference in the HR of the two samples: the mean HR value is the same for both the whole sample of candidate Type 2 (HR=0.05$\pm$0.37) and in the optically faint subsample (HR=0.07$\pm$0.33), and the hypothesis that the two distributions are actually the same can not be rejected on the basis of a KS-test (p-value=0.57).
\item The mean redshift of the whole sample, $z$=1.31$\pm$0.70, is lower than the one of the optically faint subsample (although in agreement within the errors), $z$=2.11$\pm$0.60. The hypothesis that these two redshift distributions can be obtained by by the same parent population is rejected on the basis of a KS-test (P$>$99.999\%). The difference between the redshift distributions, combined with the agreement between the HR distributions, implies that the $i>$25 sources have, on average, higher $N_H$ than those in the overall Type 2 sample (see Alexander et al. 2001; Mainieri et al. 2005).
\item Suh et al. (to be submitted) performed a multi-component modeling from far-infrared (500$\mu$m) to near-ultraviolet (2300\AA) using a 3-component SED fitting with nuclear hot dust torus, galaxy, and starburst components in order to decompose the SED into a nuclear AGN and host galaxy stellar contributions. They derived an estimate of the host galaxy stellar masses using the best-fit galaxy template, then calculating the total IR luminosities, which are integrated between 8-1000 $\mu$m from the best-fit starburst template. They then combined the infrared observations with UV observations to derive the total star formation rate (SFR), SFR$_{tot}$=SFR$_{IR}$+SFR$_{UV}$, thus estimating reliable SFRs for both obscured and unobscured sources (Arnouts et al. 2013).

The specific star formation rate (sSFR=SFR/M$^*$) distribution spans over five orders of magnitude (sSFR=[10$^{-13}$-10$^{-8}$] yr$^{-1}$) for the whole sample of candidate Type 2 AGN, while is slightly narrower for the subsample with $i>$25 (sSFR=[10$^{-11}$-10$^{-8}$] yr$^{-1}$). Moreover, the mean sSFR is almost two times larger in the optically faint subsample (3.4 $\times$10$^{-10}$ yr$^{-1}$) than in the whole sample of candidate Type 2 AGN (1.9 $\times$10$^{-10}$ yr$^{-1}$). Once again,  the hypothesis that the two distributions have been originated by the same sSFR distribution is rejected on the basis of a KS-test (P$>$99.999\%). 

\end{enumerate}

In conclusion, our data suggest that the existence of a linear trend between the hard band X-ray luminosity and X/O for Type2 AGN becomes weaker at fainter optical magnitudes, where sources have higher redshifts and the sSFR is higher, i.e., galaxies are likely in a stronger activity phase. In this subsample, the AGN contribution to the optical emission is less significant than in the X-ray, while the host-galaxy contribution is probably hidden in the optical band by the same gas responsible for the star formation. This star formation can be observed in the 3.6 $\mu$m band, where the difference in magnitude between the whole Type 2 sample (mean AB magnitude 20.5$\pm$1.1) and the $i>$25 sample (mean AB magnitude 21.8$\pm$0.7) is significantly smaller than in the $i$-band.

\subsection{Luminosity dependence of the AGN obscured fraction }\label{sec:lx_vs_obscured}

The existence of a trend between the fraction of obscured AGN and the X-ray luminosity was already shown by Lawrence \& Elvis (1982). More recently, Ueda et al. (2003) confirmed the result in the 2-10 keV (rest frame) band: at low luminosities, $L_X$$\simeq$10$^{42}$ erg s$^{-1}$, almost the whole sample is composed of obscured AGN, while unobscured sources prevail moving towards high luminosities, i.e. $L_X$$>$10$^{44}$ erg s$^{-1}$. This trend has been confirmed over the years by other works, e.g. La Franca et al. (2005), using HELLAS2XMM; Hasinger (2008), who divided the sample in unobscured and obscured sources on the basis of both the optical spectroscopic classification and X-ray absorption properties;  Ueda et al. (2014), who also found  that at higher redshifts the decline of the obscured AGN fraction starts at higher luminosities; and lastly Buchner et al. (2015). The same trend has already been confirmed by \textit{XMM-COSMOS}, on the basis of the optical classification of the sources, for both the whole survey (B10) and in different redshift bins (Merloni et al. 2014). A different result was instead found by Lusso et al. (2013), which found no clear trend with 2-10 keV luminosity of the obscured fraction of Type 1 AGN in XMM-COSMOS, using SED analysis to estimate the dust covering fraction.

\begin{figure*}[!h]
  \begin{minipage}[b]{.5\linewidth}
    \centering
   \includegraphics[width=1.0\linewidth]{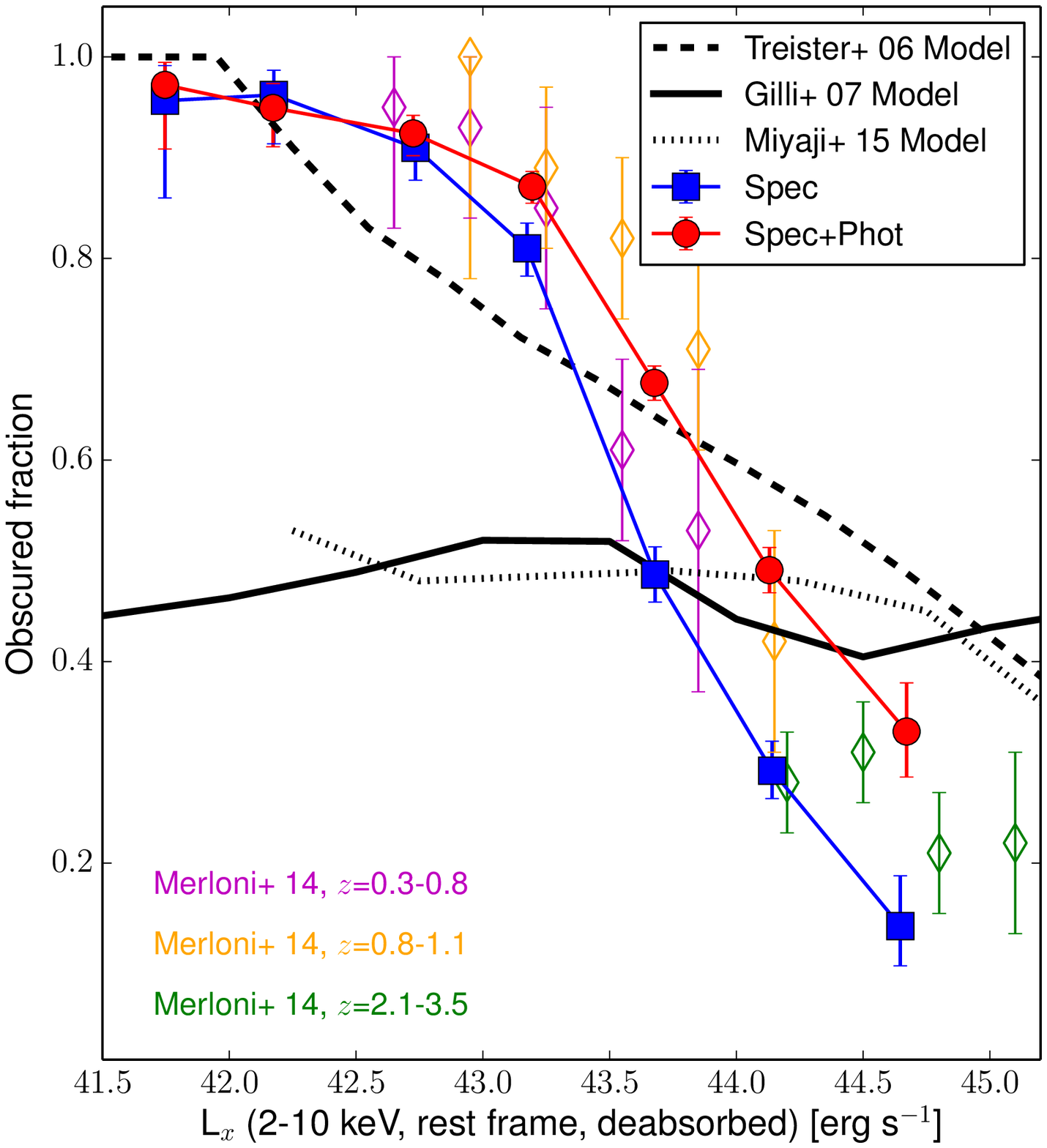}
    \label{subfig-1:i_band}
  \end{minipage}
  \hfill
  \begin{minipage}[b]{.5\linewidth}
    \centering
    \includegraphics[width=1.\linewidth]{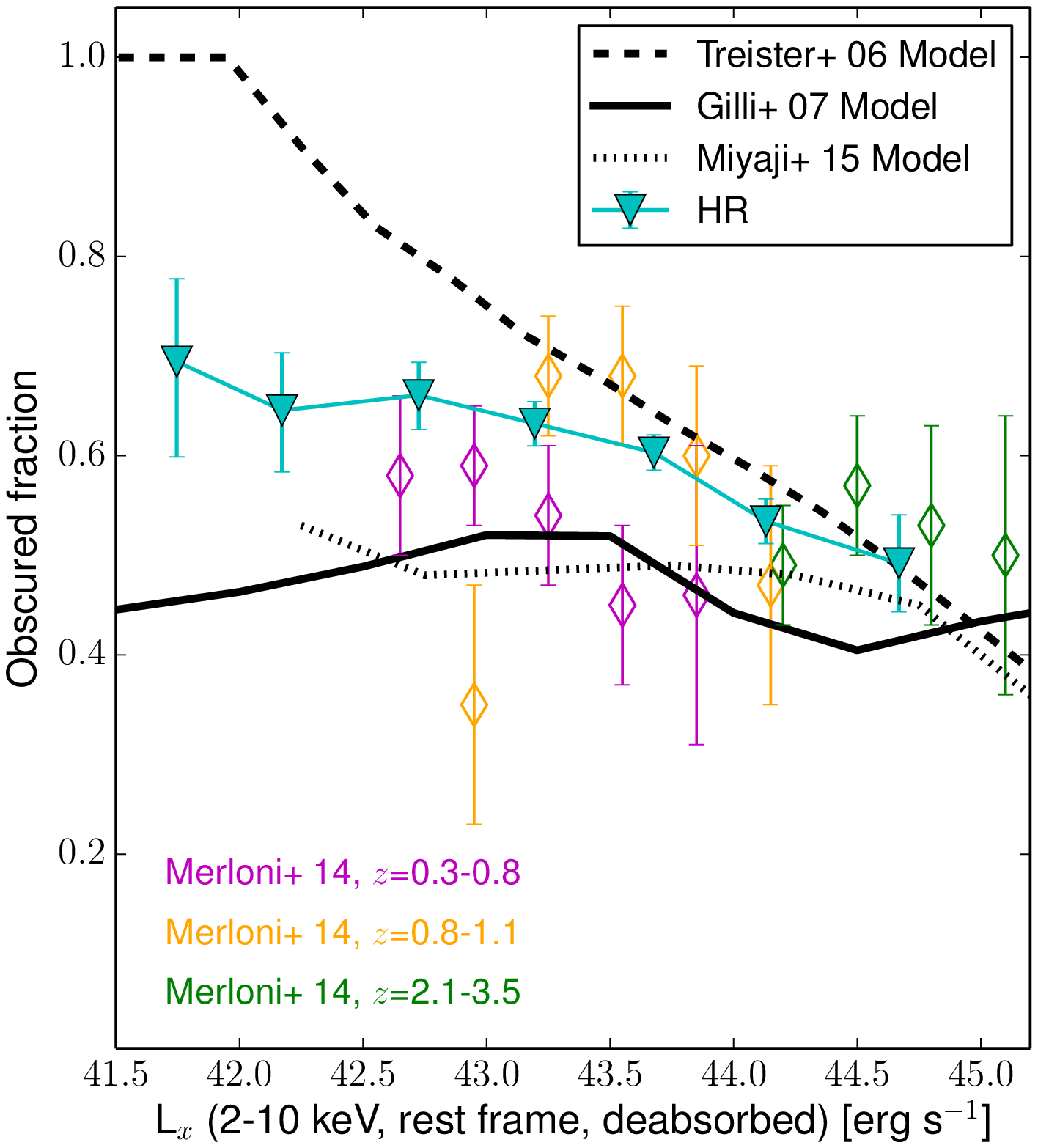}
    \label{subfig-2:k_band}
  \end{minipage}
\caption{Fraction of obscured sources as a function of 2-10 keV rest frame de-absorbed luminosity, using only spectroscopic information (blue squares, left), combined spectroscopic and photometric information (red circles, left) and X-ray only HR based information, assuming as obscured all those sources with HR$>$-0.2 (cyan triangles, right). Results obtained by Merloni et al. (2014) using subsamples of XMM-COSMOS in different bins of redshift ($z$=[0.3-0.8], magenta; $z$=[0.8-1.1], yellow; $z$=[2.1-3.5], green) are shown as diamonds. We also plot the fraction of AGN with N$_H$ $>$ 10$^{22}$ cm$^{-2}$ obtained using the XRB synthesis models by Gilli et al. (2007, solid black line), Miyaji et al. (2015) and Treister \& Urry (2006, black dashed line). All errors are 1$\sigma$ and have been calculated using Equation 26 of Gehrels (1986).}\label{fig:oscured_vs_lx}
\end{figure*}

The whole \cha \leg survey has about 900 more sources with z$>$0 and DET\_ML$>$10.8 in the hard band than \textit{XMM-COSMOS} (2354 versus $\simeq$1450), and twice better statistics than \textit{XMM-COSMOS} at luminosities lower than $L_X$$\simeq$10$^{44}$ erg s$^{-1}$ (see Figure \ref{fig:histo_lx_w_surveys}, right panel).  We studied the relation between the obscured fraction of sources versus the 2-10 keV de-absorbed luminosity: we estimated the absorption contribution following the procedure described in Section \ref{sec:nh_estimate}.  In Figure \ref{fig:oscured_vs_lx} (left panel) we plot (blue squares) the fraction of spectroscopically selected obscured AGN (i.e. the ratio between those sources which have been classified as non-BL AGN and all sources with spectroscopic type information). The whole spectroscopic type sample contains 1212 sources. 

More than 90\% of the sources at $L_X$$\leq$10$^{42}$ erg s$^{-1}$ are obscured, while the fraction of obscured sources decreases to $\simeq$80\% at $L_X$$\simeq$10$^{43}$ erg s$^{-1}$ and drops around 20\% at $L_X$$\geq$10$^{44}$ erg s$^{-1}$. 
However, there are significant uncertainties on the trend estimated using only the spectroscopic information, first of all because our spectroscopic sample is not complete (only 51.5\% of the sources included in this analysis have a spectral type) and, moreover, the selection of sources for a spectral analysis  in the COSMOS field has so far been biased towards the optically brightest sources (see Figure \ref{fig:spec_complete}), which are more likely to be unobscured broad line AGN, which could result in an under-estimation of the obscured fraction at high luminosities ($L_X$$\geq$10$^{44}$ erg s$^{-1}$). 

We therefore estimated the fraction of obscured AGN using the photometric classification for all the sources without a spectral type: the total number of sources with either a spectroscopic or a photometric type is 2343. In Figure \ref{fig:oscured_vs_lx}, left panel, we plot the fraction of obscured sources from the combined photometric and spectroscopic information in red circles: the agreement with the spectroscopic trend is good at low luminosities (more than 90\% of sources with $L_X$$<$10$^{42}$ erg s$^{-1}$ are obscured). At high luminosities ($L_X$$\geq$10$^{43.4}$ erg s$^{-1}$) the fraction of obscured sources is a factor $\simeq$2 larger, i.e $\simeq$40\%. This trend does not change significantly while computed in complete bins of redshift and luminosity.

We also compared the optical obscuration results with those obtained using the X-ray properties of the sample, using the HR (see Section \ref{sec:hr}): we divided the sources between obscured and unobscured using the HR threshold HR$_{th}$=-0.2. This is the same threshold used in B10, and it is also the average HR value at any redshift, assuming an obscured AGN spectral slope with $\Gamma$=1.4 (see Figure \ref{fig:histo_hr_vs_z_w_type}, black dotted line). The total number of sources in this third sample is 2354 (including the HR upper and lower limits), and the HR-determined obscuration fraction is plotted with cyan triangles in the right panel of Figure \ref{fig:oscured_vs_lx}. The fraction of obscured sources at low luminosities is lower than in the two previous cases ($\simeq$65\% against $\simeq$90\%, even at $L_X$$<$10$^{42}$ erg s$^{-1}$), and is comparable with the optically based result at $L_X$$\geq$10$^{44}$ erg s$^{-1}$. The discrepancy between the optical and X-ray obscured fraction at low X-ray luminosity could be due to the fact that in this luminosity range the main optical luminosity contributor is the host galaxy, the AGN being therefore hidden; conversely, in the X-rays the galaxy contribution is almost negligible at the \cha \leg fluxes and the AGN identification is unbiased (see also Merloni et al. 2014). As for the optical obscuration, the trend does not change significantly adopting complete samples in bins of redshift and luminosity.

Our results at $L_X$$>$10$^{43.5}$ erg s$^{-1}$ are also in good agreement with the fraction of obscured sources estimated using the $N_H$ value from Lanzuisi et al. (2013), where the obscured fraction of AGN is between 40 and 50\% in the 2-10 keV luminosity range $L_X$=[10$^{43.5}$-10$^{45}$] erg s$^{-1}$.

In Figure \ref{fig:oscured_vs_lx} we also compare our results with Merloni et al. (2014) \textit{XMM-COSMOS} results in different bins of redshift ($z$=[0.3-0.8], magenta diamonds; $z$=[0.8-1.1], yellow diamonds, and $z$=[2.1-3.5], green diamonds). There is a general agreement between these and our data, within the errors, using both the optical and the X-ray classification. Small differences are observed when comparing our results with theirs at $z$=[2.1-3.5], where at  $L_X$$>$10$^{44}$ erg s$^{-1}$ their results show 10-15\% more obscured sources on the basis of the optical information.

We also compare our results with the predictions of the population synthesis models of Gilli et al. (2007, black solid line) and Miyaji et al. (2015, black dotted line), both based only on the X-ray classification, with the one of and Treister \& Urry (2006, black dashed line), based on both optical and X-ray classifications. For all these models, we measured the fraction of sources with $N_H$$>$10$^{22}$ cm$^{-2}$, and folded the contribution in the two different $N_H$ ranges through the observed flux range of our survey. We divide our results in three ranges of luminosity.
\begin{enumerate}
\item At $L_X$$<$10$^{43.5}$ erg s$^{-1}$ the two models predictions diverge: the Treister \& Urry (2006) trend is more similar to the one obtained using the optical spectroscopic and photometric classifications, while our HR-based obscured fraction is closer to the predictions of the Gilli et al. (2007) and Miyaji et al. (2015) models. 
\item At 10$^{43.5}$$<$$L_X$$<$10$^{44}$ erg s$^{-1}$ there is a good agreement between the three models and our results obtained using both spectroscopic and photometric types or using the HR information. 
\item At $L_X\geq1$0$^{44}$ erg s$^{-1}$ the Treister \& Urry model overpredicts the fraction of obscured sources by 10-20\% with respect to our results using the optical classification, while the Gilli model is in good agreement with both the X-ray and optical obscuration fraction.
\end{enumerate}

On the whole luminosity range, the observed behavior on the basis of the optical classification is fairly consistent with the Treister et al. (2009) model  predictions, while the HR-based evidence of weak correlation between 2-10 keV luminosity and obscuration fraction is consistent with the Gilli et al (2007) and Miyaji et al. (2015) models predictions.

\section{Conclusions}\label{sec:conclusions}
In this paper we presented the identification procedure of optical/IR counterparts of the new 2273 \cha \leg sources. We then presented the X-ray to optical properties of the 4016 sources in the whole \cha \leg survey (i.e., the combination of the new survey and the 1743 C-COSMOS sources). The following are the main results of the identification process.

\begin{enumerate}
\item We associated the new 2773 \cha \leg point-like sources with optical/IR counterparts in three different bands ($i$, $K$ and 3.6 $\mu$m), using the likelihood ratio technique, based on both the separation between the X-ray and the optical/IR source, and the magnitude of the candidate counterpart. We found a secure counterpart in at least one of the three bands for 97\% of the X-ray sources.
\item 31 of 2273 X-ray sources have no optical/IR counterpart: even if 30-50\% of these sources could actually be spurious X-ray detections, or caused by bad optical/IR imaging, the remaining part of them are candidate obscured and/or high redshift sources.
\end{enumerate}

Thanks to the large multiwavelength coverage in the COSMOS field and to the numerous spectroscopic campaigns, we were able to provide a redshift, either spectroscopic or photometric, for almost our whole sample (96\%). We also provided a spectroscopic type and/or a photometric type from SED template best fitting.

\begin{enumerate}
\item 2151 sources of the 4016 in the whole \cha \leg survey (53.6\% of the whole sample) have a reliable spectroscopic redshift. Of these sources, 36\% are classified as BLAGN, while 59\% do not show evidence of broad lines, but only narrow emission and absorption lines. Finally, 5\% of the sources with spectroscopic information are spectroscopically identified stars.
\item We provide a photometric redshift and a related photometric classification for 3872 sources (96\%). The majority (65\%) of these sources are fitted with a non-active galaxy, even if only  a minority of sources (26\% in soft and 13\% in hard band) have $L_X$$<$10$^{42}$ erg s$^{-1}$. 9\% of the sample is fitted with an obscured AGN template and 23\% with an unobscured AGN template. Finally, 121 sources, 3\% of the whole sample, have been identified as stars on the basis of the photometric template. In XMM-COSMOS (B10) there were $\simeq$50\% Type 1 sources and $\simeq$50\% Type 2 sources: the larger fraction of obscured sources in \cha \leg is due to its flux limit three times deeper than in XMM-COSMOS. 
\item The \cha \leg luminosity distribution in the soft band peaks at $L_X\simeq$3$\times$10$^{43}$ erg s$^{-1}$ (Figure \ref{fig:histo_lx_w_surveys}, left panel), and it is an excellent bridge between deep pencil beam surveys like CDF-S (Xue et al. 2011) and large area surveys like Stripe 82 (LaMassa et al. 2013; La Massa et al. submitted). Moreover, \cha \leg covers with an excellent statistics (2285 sources in the soft band) the range of redshift 1$\leq$$z$$\leq$3, i.e. at the peak of the AGN activity and the following period (Hasinger et al. 2005). Our survey also samples with solid statistics the luminosity range below the knee of the luminosity function, up to redshift $z\simeq$4 (Figure \ref{fig:z_vs_lx}, right panel).
\end{enumerate}

Finally, we studied several X-ray-to-optical properties of our sample, especially focusing on the obscured sources.
\begin{enumerate}
\item We used the HR as a rough, purely X-ray based obscuration estimation. The mean (median) HR is HR=-0.26$\pm$0.32 (-0.30) for optically classified Type 1 sources and HR=-0.03$\pm$0.46 (-0.10) for optically classified Type 2 sources. We also studied the evolution with redshift of HR (Figure \ref{fig:histo_hr_vs_z_w_type}), and we found that, while the average HR of Type 2 sources lies above the one of Type 1 sources at any redshift, both samples show an intrinsically large dispersion. In the Type 2 sample, such a dispersion can be caused by a significant fraction of sources with a galaxy best-fit SED template being galaxy-dominated in the optical but not intrinsically obscured, therefore avoiding the possibility to correctly classify the AGN.
\item With our sample of 2798 sources in the soft band and 2363 sources in the hard band we put stronger constraints to the X-ray to optical flux ratio locus (Figure \ref{fig:x_vs_i_w_perc}). Our results confirm, with a statistics 20\% and 40\% larger in the soft and hard bands, respectively, the locus shown in C12: the new  locus is shifted to faint optical magnitudes in both soft and hard X-ray band by $\Delta$(X/O)$\simeq$0.3-0.5, without significantly changes at different fluxes.  We also studied the trend with X-ray soft flux of the $K$ (Figure \ref{fig:x_vs_k}) and 3.6 $\mu$m magnitudes and we found that the region which contains 90\% of the AGN population is considerably smaller (1.5-2 mag) than the one in  the $i$-band. This narrower relation indicates a stronger correlation of X-rays with near-infrared bands than with optical bands, a result that could be explained with a lower contribution of the nuclear extinction at near-infrared wavelengths. This last result is in general agreement with the fact that near-IR selection techniques are almost as effective as X-ray ones (Stern 2015).
\item The majority of candidate low luminosity AGN or non active galaxies (i.e. sources with $L_X$$<$10$^{42}$ erg s$^{-1}$) have low X-ray fluxes and bright optical magnitudes (Figure \ref{fig:x_vs_i_w_perc}). However, there is a fraction of sources with low $L_X$ which shows X-ray to optical properties consistent with those of sources with $L_X$$>$10$^{42}$ erg s$^{-1}$: 15\% and 28\% of sources with  $L_X$$<$10$^{42}$ erg s$^{-1}$ lie inside the \cha \leg X-ray to optical flux ratio locus in the soft and hard bands, respectively. The fraction is considerably higher in the hard band, where it is more likely to observe obscured AGN.
\item We confirm the existence of a correlation between X/O and the luminosity in the 2-10 keV band for Type 2 sources (Figure \ref{fig:xo_vs_lx}). We also confirm that at faint magnitudes ($i>$25) the trend is weaker, and our data suggest that this happens at higher redshifts, where the sSFR is higher and the AGN contribution to the optical emission is less significant than the one in the X-ray.
\item We extend to low luminosities the well known, inverse correlation between the fraction of obscured AGN and the hard band luminosity: the fraction of optically classified obscured AGN is of the order of 90\% at $L_X$$<$10$^{42}$ erg s$^{-1}$ and drops to $\simeq$40\% at $L_X$$>$10$^{43.5}$ erg s$^{-1}$. The observed behavior is fairly consistent with the Treister et al. (2009) AGN synthesis model  predictions. On the other hand, if an X-ray classification criterion based on the HR is adopted, the lack of a strong correlation between obscured fraction and luminosity is consistent with the Gilli et al (2007) and Miyaji et al. (2015) models predictions.  A higher spectroscopic completeness, coupled with a proper X-ray spectral analysis would be needed to fully capture the dependence on luminosity of the obscured AGN fraction.
\end{enumerate}

\section{Acknowledgments}
This research has made use of data obtained from the \cha Data Archive and software provided by the \cha X-ray Center (CXC) in the CIAO application package.

This work was supported in part by NASA Chandra grant number GO7-8136A (F.C., S.M., V.A., M.E.); PRIN-INAF 2014 "Windy Black Holes combing galaxy evolution" (A.C., M.B., G.L. and C.V.); the FP7 Career Integration Grant ``eEASy'': ``Supermassive black holes through cosmic time: from current surveys to eROSITA-Euclid Synergies" (CIG 321913; M.B. and G.L.); the Collaborative Research Council 956, sub-project A1, funded by the Deutsche Forschungsgemeinschaft (DFG; A. K.); UNAM-DGAPA Grant PAPIIT IN104113 and  CONACyT Grant Cient\'ifica B\'asica \#179662 (T.M.); the Greek General Secretariat of Research and Technology in the framework of the programme Support of Postdoctoral Researchers (P.R.); NASA award NNX15AE61G (R.G.);  the Swiss National Science Foundation Grant PP00P2\_138979/1 (K.S.); the World Premier International Research Center Initiative (WPI Initiative), MEXT, Japan (J.D.S.); the Center of Excellence in Astrophysics and Associated Technologies (PFB 06), by the FONDECYT regular grant 1120061 and by the CONICYT Anillo project ACT1101 (E.T.); the European Union's Seventh Framework programme under grant agreements 337595 (ERC Starting Grant, ``CoSMass'') and 333654 (CIG, AGN feedback; V.S.). B.T. is a Zwicky Fellow.

\end{document}